\documentclass{elsarticle}
\pdfoutput=1
\usepackage{hyperref}
\usepackage{aas_macros}

\journal{Planetary and Space Science}

\biboptions{authoryear}

\begin{document}

\begin{frontmatter}

\title{The mineralogy of newly formed dust in active galactic nuclei}
\author[affil1]{Sundar Srinivasan\corref{cor1}}
\ead{sundar@asiaa.sinica.edu.tw}
\author[affil1]{F. Kemper}
\ead{ciska@asiaa.sinica.edu.tw}
\author[affil2]{Yeyan Zhou}
\ead{201111161006@mail.bnu.edu.cn}
\author[affil2]{Lei Hao}
\ead{haol@shao.ac.cn}
\author[affil3,affil4]{Sarah C. Gallagher}
\ead{sgalla4@uwo.ca}
\author[affil5,affil6]{Jinyi Shangguan}
\ead{shangguan@pku.edu.cn}
\author[affil5,affil6]{Luis C. Ho}
\ead{lho.pku@gmail.com}
\author[affil5]{Yanxia Xie}
\ead{yanxia.ts@gmail.com}
\author[affil1]{Peter Scicluna}
\ead{peterscicluna@asiaa.sinica.edu.tw}
\author[affil8,affil7,affil1]{Sebastien Foucaud}
\ead{sebitosecos@gmail.com}
\author[affil1,affil7]{Rita H.T. Peng}
\ead{htpeng1108@gmail.com}
\cortext[cor1]{Principal corresponding author}
\address[affil1]{Institute of Astronomy \& Astrophysics, Academia Sinica, 11F, Astronomy-Mathematics Building, No. 1, Roosevelt Rd, Sec 4, Taipei 10617, Taiwan, Republic of China}
\address[affil2]{Shanghai Astronomical Observatory, Chinese Academy of Sciences, 80 Nandan Road, Shanghai 200030, China}
\address[affil3]{Department of Physics and Astronomy, University of Western Ontario, London, ON N6A 3K7, Canada}
\address[affil4]{Centre for Planetary and Space Exploration, London, ON N6A 3K7, Canada}
\address[affil5]{Kavli Institute for Astronomy and Astrophysics, Peking University, Beijing 100871, China}
\address[affil6]{Department of Astronomy, School of Physics, Peking University, Beijing 100871, China}
\address[affil7]{Department of Earth Sciences, National Taiwan Normal University, Taipei 11677, Taiwan, Republic of China}
\address[affil8]{Center for Astronomy and Astrophysics, Shanghai Jiao Tong University, Shanghai 200340, China}
\begin{abstract}
The tori around active galactic nuclei (AGN) are potential formation sites for large amounts of dust, and they may help resolve the so-called dust budget crisis at high redshift. We investigate the dust composition in 53 of the 87 Palomar Green (PG) quasars showing the 9.7  $\mu$m silicate feature in emission. By simultaneously fitting the mid-infrared spectroscopic features and the underlying continuum, we estimate the mass fraction in various amorphous and crystalline dust species. We find that the dust consists predominantly of alumina and amorphous silicates, with a small fraction in crystalline form. The mean crystallinity is 8$\pm$6\%, with more than half of the crystallinities greater than 5\%, well above the upper limit determined for the Galaxy. Higher values of crystallinity are found for higher oxide fractions and for more luminous sources.\\
\end{abstract}


\end{frontmatter}


\section{Dust formation in active galactic nuclei}
The observational appearance of quasars, particulary the obscuration of the central source, has been explained
by the Unified Model, which invokes the presence of a dusty torus \citep[e.g.][]{Antonucci_93_Unified,Bianchi_12_AGN,Netzer_15_Revisiting,Mason_15_Dust}.
Although initially envisioned as a static entity, the torus is now thought to be dynamic, and represents either an infalling gas \citep{Hopkins_12_origins} 
or an outflow of gas \citep[e.g.,][]{KoeniglKartje1994,ElitzurShlosman2006,Keating+2012} with dust embedded in it. Silicates were proposed as the main constituent
of this dust \citep{Stenholm_94_Silicate}. Increasingly sophisticated radiative transfer models of the dusty tori predicted the appearance
of both silicate absorption and emission, depending on the viewing angle \citep{PierKrolik1992,Nenkova_02_Dust,vanBemmel_03_New,Hoenig_06_Radiative,Fritz_06_Revisiting,Stalevski_12_3D}, 
although initially only silicate absorption features
were observed, indicative of a near edge-on viewing angle.
The detection of silicate emission in active galactic nuclei (AGNs) is considered evidence for the unified model \citep{Siebenmorgenetal2005,Haoetal2005,Sturmetal2005}. In addition to the presence of silicates, other minerals have been 
predicted to condense as well, analogous to the dust formation sequence in evolved stars \citep{Elvisetal2002}. In principle,
the exact condensation sequence, which can be probed by determining the dust composition, reveals information about the physical
conditions in the dust forming (or processing) environments. Nevertheless, most infrared spectroscopic studies have focussed 
on measuring the optical depth in the silicate emission or absorption feature, using it as a proxy for the column density along the
line-of-sight \citep[e.g][]{Shietal2006,Hao_07_Distributiona,Sales_11_Compton,Hatziminaoglouetal2015}. Some studies have even spatially explored variations in the column density, with particular emphasis on the nearby AGN NGC 1068 
\citep[e.g.][]{Rhee_06_First,Mason_06_Spatially,Poncelet_06_new,Lopez-Gonzaga_14_Revealing,Alonso-Herrero_16_mid,Lopez-Rodriguez_16_Mid}, using
standard astronomical silicate opacities such as those published by \citet{Draine_07_Infrared}. While it was already reported by \citet{Sturmetal2005} that the 
silicates seen towards quasars differ from the Galactic interstellar silicates observed toward the Galactic Centre \citep[{e.g.},][]{Kemperetal2004}, only a handful studies have since attempted to explain the difference in spectral appearance in a relatively small
number of objects, 
using a different dust composition \citep{Jaffe_04_central,Markwick-Kemperetal2007,KohlerLi2010}, variations in the silicate mineralogy \citep{Xieetal2014,Xie_15_Tale}, different grain properties \citep{Li_08_anomalous,Smithetal2010}
or optical depth effects \citep{Nikuttaetal2009}. 
In this paper, we present the first results of a systematic study of AGN dust mineralogy by fitting the mid-infrared spectrum of a sample of Palomar Green (PG) quasars showing silicate emission following the method described by \citet{Markwick-Kemperetal2007}, and trying to correlate the results with the physical properties of the AGN. We describe our sample selection in Section \ref{sec:data}, and our model and fitting procedure in Section \ref{sec:analysis}. We show our results in Section \ref{sec:results} and discuss the implications of these results in Section \ref{sec:discussion}.

\section{Sample selection}
\label{sec:data}
We derive our sample from the 87 nearby ($z \leq 0.5$) optically selected luminous broad-line QSOs in the Palomar Bright Quasar Survey (BQS) Catalog \citep[][hereafter, the Palomar-Green or PG sample]{SchmidtGreen1983}. These low-redshift objects were famously studied by \citet{BorosonGreen1992}, combining data at X-ray, optical, and radio wavelengths. More recently, the mid-infrared spectroscopic features of these objects (in particular, the strength of the 9.7 $\mu$m silicate emission) were studied by \citet{Shietal2006} and \citet{Hatziminaoglouetal2015}. \citet{Shietal2014} obtained {\em Spitzer} photometry and spectroscopy for the entire PG sample. \citet{Petricetal2015} combined near-, mid-, and far-infrared information and computed the rest-frame luminosity and cold dust content for 85 of these 87 quasars.\\

Inclusion in the sample discussed by \citet{Petricetal2015} ensures a {\em Herschel} PACS 70 $\mu$m photometric measurement, which is essential to estimate the continuum emission longward of the 18 $\mu$m feature. \citet{Shietal2014} also provide MIPS 70 $\mu$m photometry which can be used for the same purpose. By carefully re-reducing the \citet{Petricetal2015} {\em Herschel} data, Shangguan et al. ({\em in prep.}) derive lower 70 $\mu$m fluxes (15\% lower on average). In this paper, we use the Shangguan et al. determinations for the 85 PG quasars studied by \citet{Petricetal2015}. For the two remaining objects, we use the MIPS 70 data from \citet{Shietal2014}. This is reasonable, as the MIPS fluxes are in general agreement with the PACS values.\\

For each source in the PG sample, we obtained archival spectra from the \emph{Combined Atlas of Sources with Spitzer IRS Spectra\footnote{\url{http://cassis.sirtf.com}}} \citep[CASSIS;][]{Lebouteilleretal2011,Lebouteilleretal2015}. For each source, we collect all available low-resolution spectra from the CASSIS database. If more than one observation exists for each module (short-low, 5-14 $\mu$m, and long-low, 14-40 $\mu$m), we average these observations. We then account for any mismatch between the short-low and long-low modules by scaling the former to the latter. Finally, we scale the synthetic photometry in the MIPS 24 band computed for each spectrum to the MIPS 24 flux as measured by \citet{Shietal2014}.\\

The spectra show a great variety in appearance, and many of them were included in the Spitzer Quasar and ULIRG Evolution Study \citep[QUEST;][]{Schweitzeretal2006,Netzeretal2007,Veilleuxetal2009}, where their spectral features and long-wavelength emission were discussed in detail. We briefly descibe some common properties of the spectra here. Many of the spectra show emission features due to polycyclic aromatic hydrocarbons (PAHs) at rest wavelengths of 6.2, 7.7, 8.6, 11.2, and 12.7 $\mu$m, probably associated with star formation activity in the host galaxy. Furthermore, the H$_2$ S(3) 9.65 $\mu$m line can often be seen, as well as atomic emission lines related to starburst activity, in the form of [Ar{\sc ii}] 6.986 $\mu$m, [Ne{\sc ii}] 12.81 $\mu$m, and [Ne{\sc iii}] 15.56 $\mu$m lines.  The spectra also feature some high ionization-potential lines typically found in AGN, such as the 10.51 $\mu$m [Si{\sc iv}], the [Ne{\sc v}] 14.32 $\mu$m and the [O{\sc v}] 25.89 $\mu$m lines. The dust emission spectrum shows a range of appearances of the solid state features. The amorphous silicate features at $\sim$10 and 18 $\mu$m are evident, and many spectra also show evidence of crystalline silicate as relatively narrow features at $\sim$19, 23, and 27.5 $\mu$m.\\

We removed PG~1352+182 from our sample, as its spectrum was too noisy and its short-low spectrum was too faint compared to the long-low one. For the remaining sources, we collect the relevant physical parameters from the literature. We use the redshifts determined by \citet{BorosonGreen1992} to shift the wavelengths to the rest frame. In addition, we also collect the absolute $B$ magnitude from the BQS paper \citep{SchmidtGreen1983}. The PG sample is one of the most thoroughly explored AGN/QSO datasets. AGN luminosities and black hole masses for the sample have been systematically explored by many authors \citep[][]{BaskinLaor2005,VestergaardPeterson2006,Geetal2016}. We use the values from \citet{Geetal2016} for the host-corrected black hole mass and accretion rate. The properties of our sample are summarised in Table \ref{tab:sample}. 

\begin{table}[tbh]
\begin{center}
\footnotesize
\caption[]{The sample studied in this paper. The PG name and SIMBAD name are shown in the first two columns, followed by the positions. The redshifts are from \citet{BorosonGreen1992}. Absolute $B$ magnitudes are taken from \citet{SchmidtGreen1983}. The last two columns are the host-corrected black hole masses and Eddington ratios from \citet{Geetal2016}. Table \ref{tab:sample} is published in its entirety in machine-readable format. The first five rows are shown here for guidance regarding its form and content.}\label{tab:sample}
\begin{tabular}{llllllll}
\hline\noalign {\smallskip}
PG name & SIMBAD& RA & Dec & $z$ & M$_{\rm B,abs}$ & $\frac{M_{\rm BH}^{\rm corr}}{M_\odot}$ & $\frac{L_{\rm bol}}{L_{\rm Edd}}$\\
 & & & & & & ($\log$) & ($\log$)\\
\hline\hline
0003+158      & PG 0003+158   & 0 5 59.24     & 16 9 49.0     & 0.451   & --26.38  & 9.27  & --0.388  \\
0003+199      & MRK 0335      & 0 6 19.58     & 20 12 10.6    & 0.0260  & --22.14  & 7.15  & --0.566  \\
0007+106      & MRK 1501      & 0 10 31.01    & 10 58 29.5    & 0.0890  & --22.56  & 8.69  & --1.09   \\
0026+129      & PG 0026+129   & 0 29 13.70    & 13 16 3.89    & 0.142   & --24.76  & 8.59  & --0.818  \\
0043+039      & PG 0043+039   & 0 45 47.23    & 4 10 23.4     & 0.385   & --26.09  & 9.13  & --0.722  \\
\hline
\end{tabular}
\end{center}
\end{table}
\normalsize

\section{Analysis}
\label{sec:analysis}
In our earlier work \citep{Markwick-Kemperetal2007}, we developed a mineralogy model to determine the dust composition. In that paper, the continuum was computed using a spline fit, and the continuum-subtracted spectrum was modeled as a linear combination of a number of dust species. We modify this procedure by assuming a power-law continuum in the mid-infrared range. This parametric form allows us to quantify our treatment of the continuum, and these parameters can be solved for simultaneously with those for the dust features. These simplifications to the physics allow us to focus on the details of the dust mineralogy in this work. By assuming that the carriers of the feature and the continuum have the same temperature, we are able to decouple the mineralogy from the overall continuum in the form of a continuum-divided spectrum, reducing the number of free parameters in the problem. Our assumptions are valid since the medium is entirely optically thin at the relevant wavelengths.\\

\begin{figure*}
\includegraphics[width=4in]{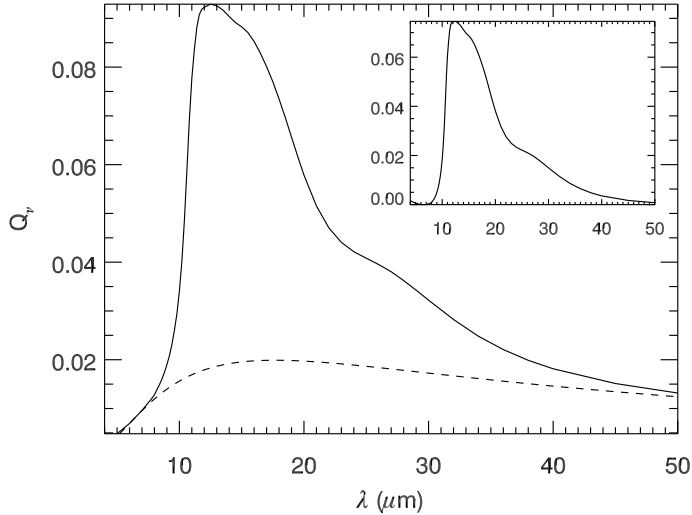}
\caption{The absorption efficiency $Q_\nu$ for corundum, with a robust polynomial fit (dashed line) to represent the underlying continuum, which is be used to compute the continuum-subtracted $Q_\nu$ ({\em inset}).}
\label{fig:Qsub}
\end{figure*}

The mid-infrared spectrum is modeled as a linear combination of a number of dust species overlaid onto a power-law continuum with spectral index $\alpha$:

\begin{equation}
F_{\nu,{\rm mod}}=A\lambda^\alpha\left(1+\sum_{j=1}^{N}c_jQ_{\nu,j}\right),
\label{eqn:model}
\end{equation}
with $N$ the total number of dust species considered, and $Q_{\nu,j}$ and $c_j$ the (continuum-subtracted) absorption efficiency and relative number of dust grains of the $j^{\rm th}$ species respectively. In this work, we fit the spectra using a combination of amorphous and crystalline dust species, as well as interstellar PAHs. According to the classical O--rich dust condensation \citep{Tielens1990,LoddersFegley1999,GailSedlmayr1999}, we expect the formation of refractory oxides (alumina, periclase) at higher temperatures, followed by forsterite and eventually olivine, as the Mg and Si condense into silicates with lowering temperature and increased density. In this study, we fit the mid-IR spectra using the oxides alumina and periclase, olivine and magnesium-rich olivine for amorphous silicates, and forsterite and clinoenstatite for crystalline silicates. We used the average interstellar PAH profile derived by \citet{Hony_01_CH} to fit the PAH emission\footnote{A more detailed treatment of the PAH features is beyond the scope of this paper.}. The details of the dust species are shown in Table \ref{tab:dustdetails}. Each species is assumed to consist of a continuous distribution of ellipsoids (CDE, \citealt{BohrenHuffman}) of fixed volume, corresponding to a radius of 0.1  $\mu$m for spherical grains. In each case, we compute the continuum-subtracted absorption efficiency $Q_\nu$ using a robust polynomial fit to the continuum under the features, using wavelengths in the 4--8 $\mu$m range and the 35--50 $\mu$m range. Figure \ref{fig:Qsub} shows an example using the absorption efficiency for corundum computed from the optical constants of \citet{Begemannetal1997}.\\

We use the robust non-linear least squares curve fitting routine {\tt MPFIT} \citep{Markwardt2009} in IDL to compute the best-fit to each spectrum. We fit the model given by Equation \ref{eqn:model} between the wavelengths 8 and 35 $\mu$m, constraining the continuum beyond the IRS spectrum using the Herschel PACS (or MIPS 70) $\mu$m photometery where available. Due to the robust nature of the {\tt MPFIT} routine, the best fit is not affected by sharp atomic features that are neither accounted for in the model, nor blocked in the data. We use the densities (Table \ref{tab:dustdetails}) to convert the best-fit weights into mass fractions for each grain species. We also compute the crystalline fraction, defined as the ratio of the mass fraction in crystalline silicates (forsterite + enstatite) to the total mass fraction in silicates (crystalline + amorphous).\\

\begin{table}[tbh]
\begin{center}
\footnotesize
\caption[]{Details of the dust species used in this study. For each grain species, we compute the $Q$ values assuming CDE with a volume corresponding to a spherical grain of 0.1  $\mu$m. The PAH emission is fit with an average interstellar profile. The material densities are taken from \citet{KleinDutrow2008}.}\label{tab:dustdetails}
\begin{tabular}{lll}
\hline\noalign{\smallskip}
Species & Grain density & Optical constants\\
 & (g cm$^{-3}$) & \\
\hline\hline
{\em Amorphous} & & \\
\hspace{0.5cm} Corundum & 4.02 & \citet{Begemannetal1997}\\
\hspace{0.5cm} Periclase & 3.56 & \citet{Hofmeisteretal2003}\\
\hspace{0.5cm} Olivine & 3.79 & \citet{Dorschneretal1995}\\
\hspace{0.5cm} Mg-rich olivine & 3.22 & \citet{Jaegeretal2003}\\
{\em Crystalline} & & \\
\hspace{0.5cm} Forsterite & 3.2 & \citet{Jaegeretal1998}\\
\hspace{0.5cm} Clinoenstatite & 3.28 & \citet{Jaegeretal1998}\\
{\em PAHs} & -- & \citet{Hony_01_CH}\\
\hline
\end{tabular}
\end{center}
\end{table}
\normalsize

\begin{table}[tbh]
\begin{center}
\footnotesize
\caption[]{Best-fit abundance fractions for each dust species for the first five sources in our sample. The first five rows are shown here for guidance regarding its form and content. The entire table, including the uncertainties in the parameters, is published in machine-readable format.\label{tab:results}}
\begin{tabular}{lllllllll}
\hline\noalign{\smallskip}
PG name & \multicolumn{6}{c}{Mass fractions (\%)} & Crys. & Comment\\
& Al$_2$O$_3$ & MgO & Olivine & Mg-rich & Forsterite & Clino- & (\%) & \\
 & & & & olivine & & enstatite & & \\
\hline
0003+158 & 46.4 & 8.29 & 41.2 & 0.00 &  2.53 & 1.53 & 8.95 & S/N $<$ 20\\
0003+199 & 68.1 & 0.00 &  30.0 & 0.00 &  1.85 & 0.00 &  5.81 & --\\
0007+106 & 0.00 &  39.5 & 59.7 & 0.00 &  0.794 & 0.00 &  1.31 & --\\
0026+129 & 43.0 & 18.7 & 35.8 & 0.00 &  2.49 & 0.00 &  6.49 & --\\
0043+039 & 32.3 & 10.8 & 18.8 & 27.5 & 4.43 & 6.19 & 18.6 & S/N $<$ 20\\
\hline
\end{tabular}
\end{center}
\end{table}
\normalsize


\section{Results: the mineralogy of quasars in the PG sample}
\label{sec:results}
\begin{figure*}
\includegraphics[width=4in]{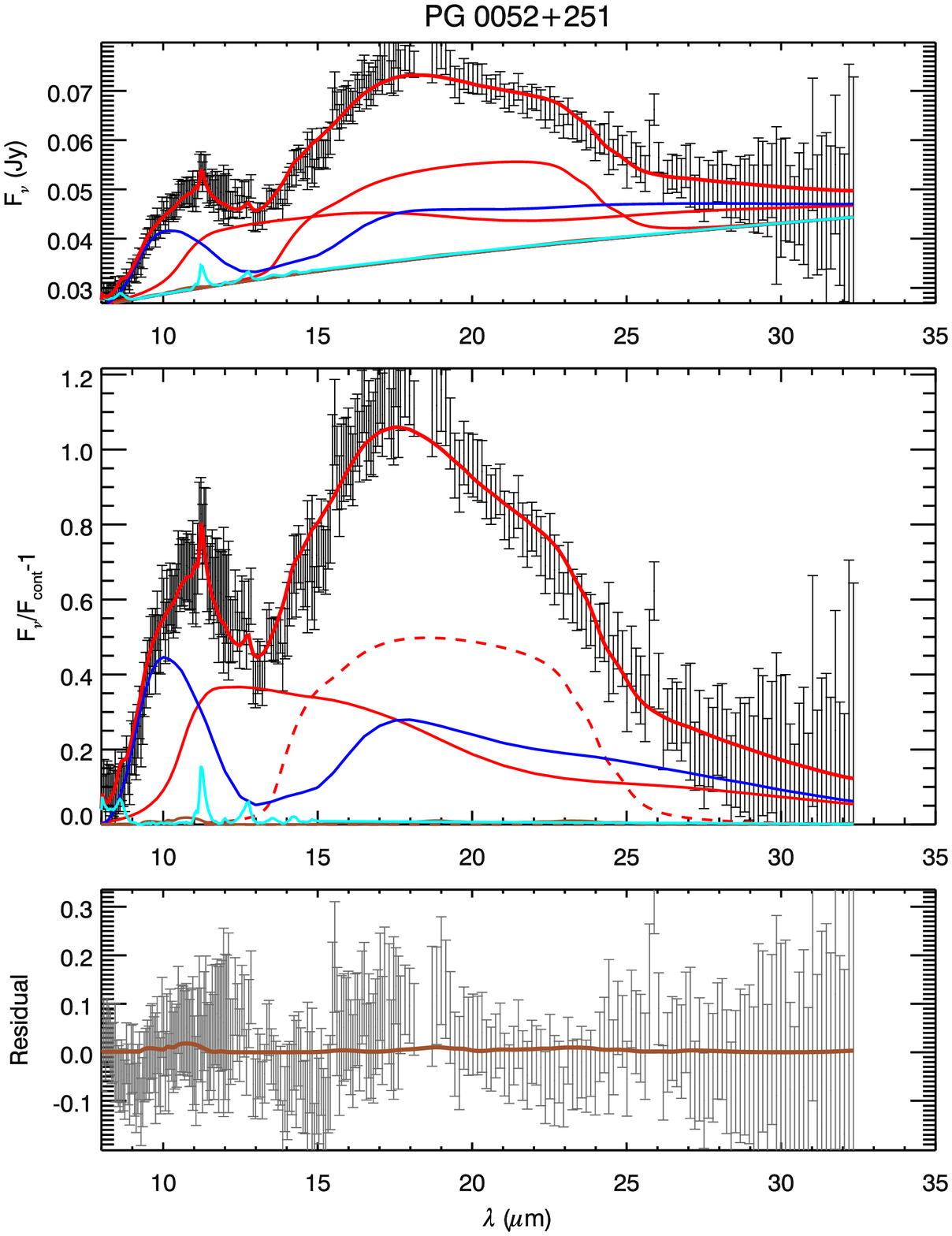}
\caption{Comparison of the IRS spectrum for PG~0052+251 (black) with the best-fit model for the source. Each panel shows the full model (thick red) as well as the indivudual components (corundum: solid red, periclase: dashed red, olivine: solid blue, Mg-rich olivine: dashed blue, forsterite: solid brown, clinoenstatite: dashed brown, PAH: cyan), depending on their fit weights. {\em Top}: original spectrum; {\em Middle}: continuum-subtracted; {\em Bottom}: Residual spectrum after all amorphous components and PAHs are subtracted, fit with the crystalline component (brown). The fit result for this spectrum shows that it is composed mostly of oxides and amorphous silicates, with some PAH emission.
\label{fig:results1}}
\end{figure*}

\begin{figure*}
\includegraphics[width=4in]{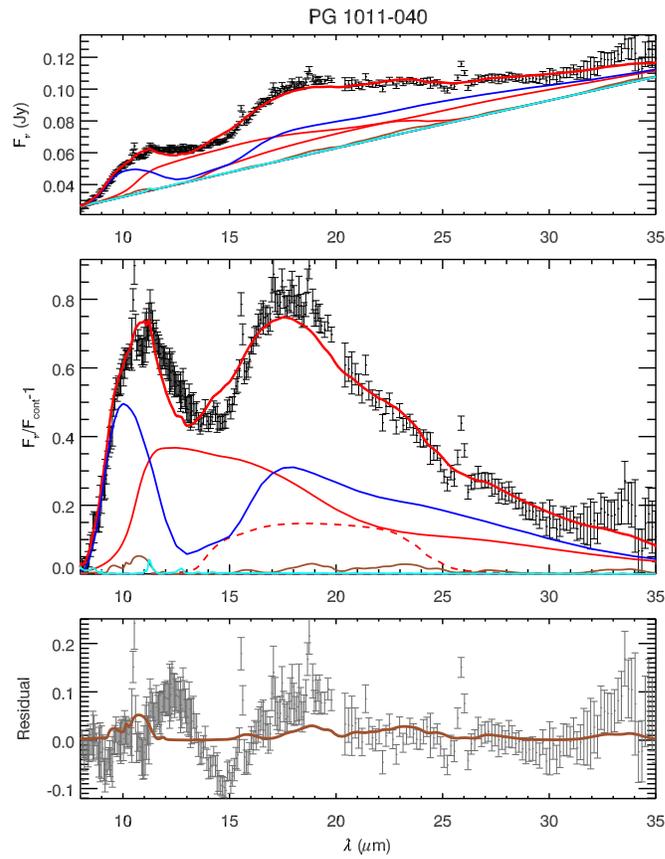}
\caption{Fit results for PG~1011--040, showing mostly amorphous components, with perhaps a small amount of forsterite. Symbols same as in Figure \ref{fig:results1}.
\label{fig:results2}}
\end{figure*}

\begin{figure*}
\includegraphics[width=4in]{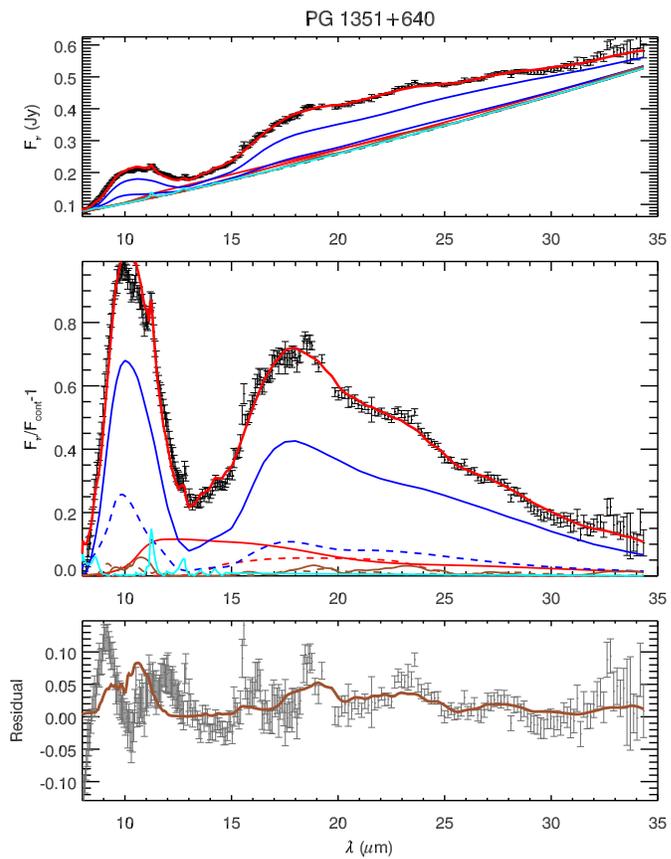}
\caption{Fit results for PG~1351+640. The very good quality spectrum also has a very good quality overall fit, with olivine dominating the fit. Symbols same as in Figure \ref{fig:results1}.
\label{fig:results3}}
\end{figure*}

\begin{figure*}
\includegraphics[width=4in]{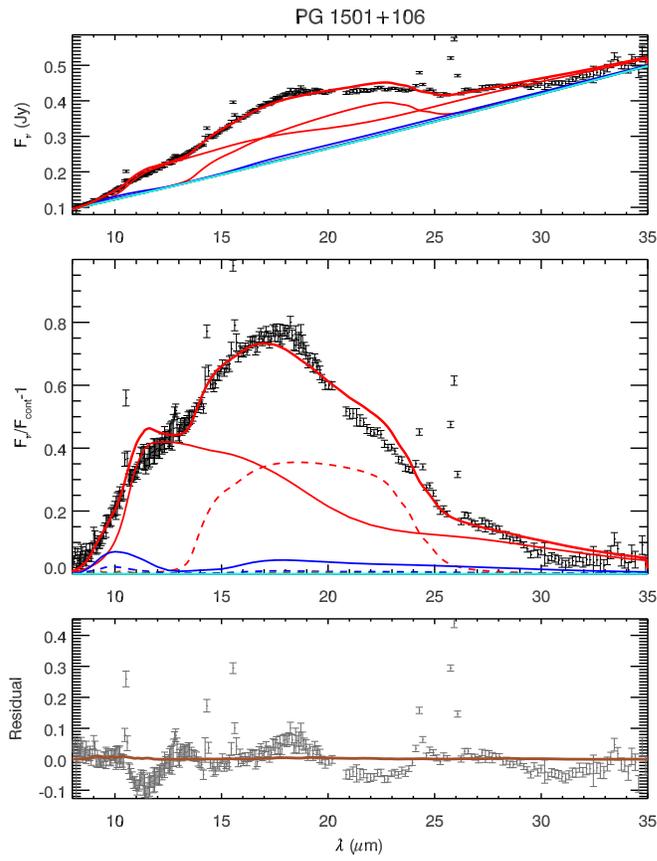}
\caption{Fit results for PG~1501+106, implying a large oxide fraction. Symbols same as in Figure \ref{fig:results1}.
\label{fig:results4}}
\end{figure*}

\begin{figure*}
\includegraphics[width=4in]{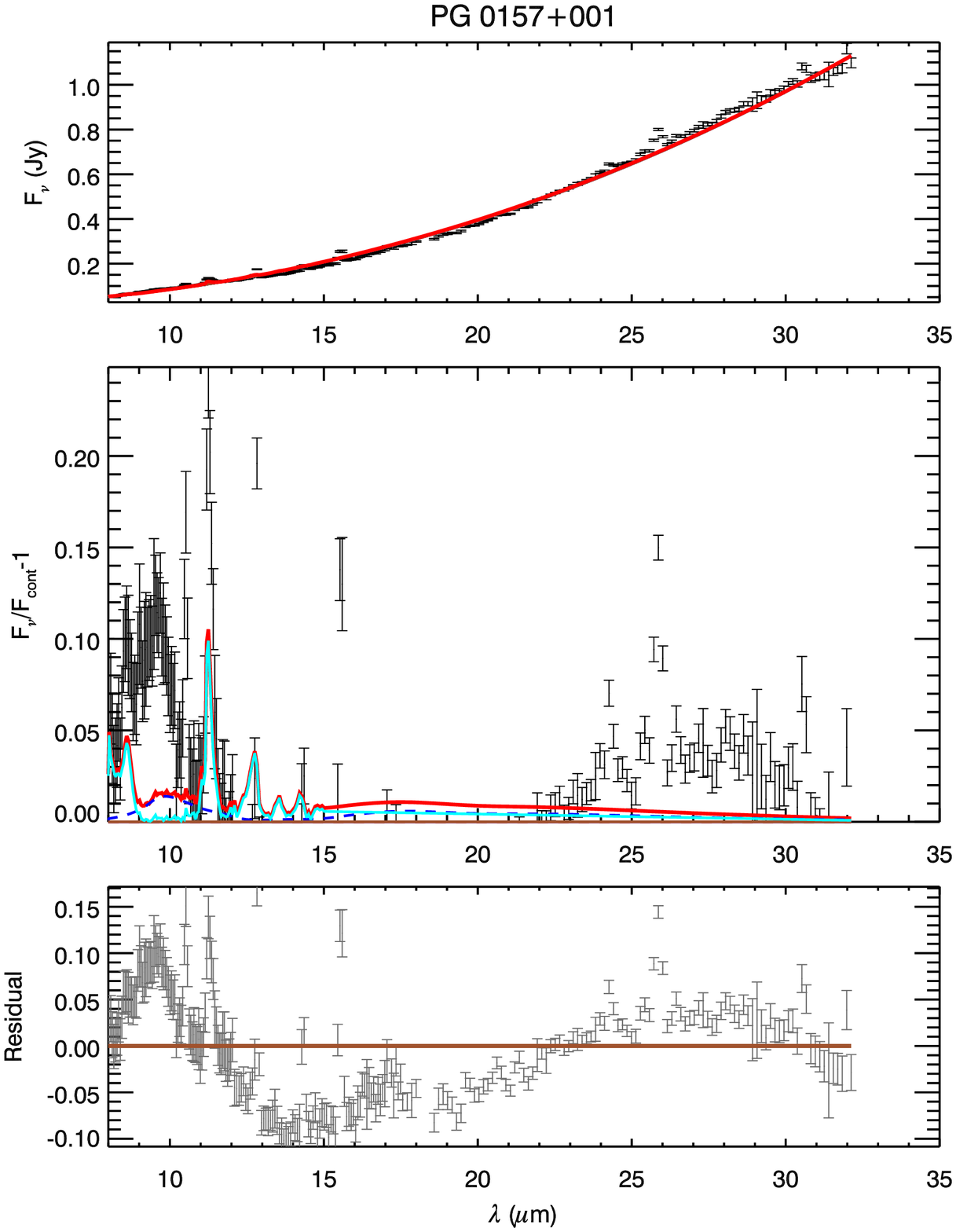}
\caption{Fit results for PG~0157+001. Symbols same as in Figure \ref{fig:results1}. In this case, continuum over-subtraction results in an underestimate of the (already low) silicate contribution. The subtracted spectrum is dominated by PAHs.
\label{fig:results5}}
\end{figure*}

We find good overall agreement between the best-fit models and the spectra -- the fits are typically robust against noisy parts of the spectrum, reproducing the global features. Figures \ref{fig:results1}--\ref{fig:results4} show some examples of our fit results. In each figure we show the total fit, the continuum-subtracted fit, and the crystalline component fit compared to the residual spectrum after all amorphous model components have been subtracted. Figures \ref{fig:results1}--\ref{fig:results4} are examples of good overall fits to spectra showing that, in general, the dust in the PG sample is composed mostly of amorphous oxides and silicates, with a small fraction of crystalline silicates. While the overall nature of the spectrum is modelled well, the contributions from the individual dust species are unable to reproduce the correct shape of the silicate features. There are three sources which give rise to these disagreements. In some cases, the continuum is not accurately estimated. This is especially the case with spectra showing very weak silicate emission -- an overestimate of the continuum sometimes also results in an underestimate of the silicate fraction in the dust. An example is shown in Figure \ref{fig:results5}, which shows the fit to PG~0157+001. The continuum-subtracted spectrum displays strong PAH emission with weak silicate features that are not reproduced well by the fit. Another reason is that the amorphous components are not able to account for the full range of broad features in the spectra. This can be reconciled by considering a larger range of grain properties. In this paper, the amorphous component is represented by the combination of corundum, periclase, and olivines with a fixed set of dust properties -- CDE grains with fixed volume corresponding to 0.1 $\mu$m spheres. Exploring the change in the shape of the broad features by varying these dust properties can help improve the fit to these data. We are improving our model treatment by exploring these dust parameters. Finally, many of the spectra have low signal-to-noise, which affects the best-fit estimate. We do not consider spectra with median S/N less than 20 (23 objects) in the subsequent discussion. We also exclude PG~1126--041, which has a very noisy short-low spectrum, and nine other sources featuring PAH-dominated spectra and/or continuum over-subtraction leading to large overestimates for the crystallinity, leaving 53 sources. The comments column in Table \ref{tab:results} identifies all sources that are excluded from further discussion.\\

We convert the best-fit relative weights of the amorphous and crystalline dust components into mass fractions using the densities from Table \ref{tab:dustdetails}. These mass fractions are available in Table \ref{tab:results}. In addition, we compute the mass fraction of silicates in crystalline form by dividing the mass fraction of forsterite and clinoenstatite by the total mass fraction in amorphous and crystalline silicates. This crystallinity is also shown in the table, and it can be used to investigate how the dust formation or processing in the dusty regions around AGN differs from that in other environments.\\

\begin{figure}
\includegraphics[width=\textwidth]{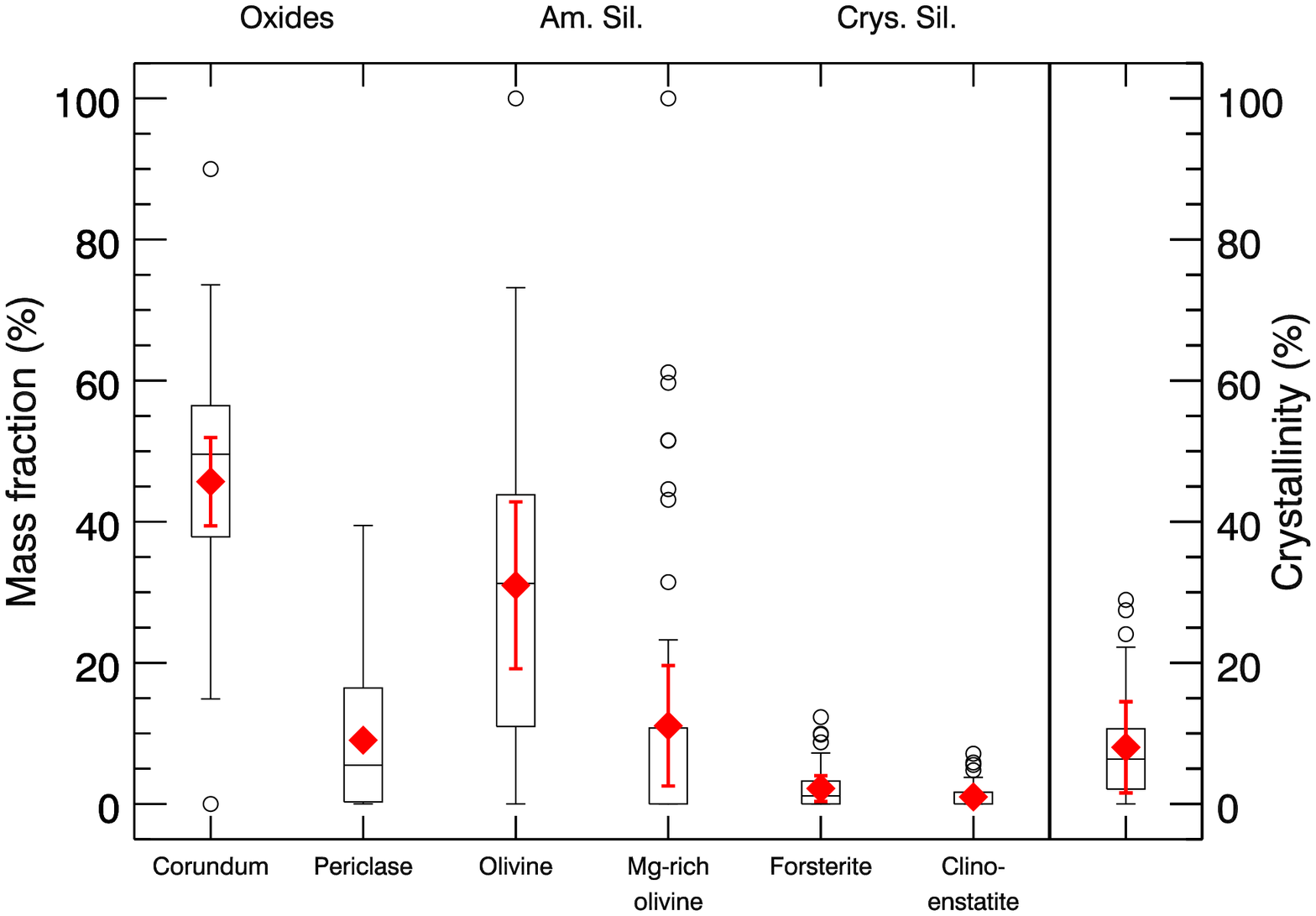}
\caption{Mass fractions of various dust species for the sample of 53 quasars studied. The boxes indicate the interquartile range, with the thick horizontal line marking the median value, and the diamonds the mean value. The whiskers extend out to 1.5 times the box length in either direction of the median, and outliers are shown as open circles.
\label{fig:boxplot}}
\end{figure}

Figure \ref{fig:boxplot} summarises our results for the 53 quasars with high S/N spectra in the form of a box-whisker plot for the relative mass fractions of these dust components, as well as the crystallinity. The box width is the interquartile range, enclosing 50\% of the central values, with the median marked as a thick  horizontal line. The whiskers extend out to 1.5 times the quartile in each direction. For each species, the mean value is also shown as a red diamond. We find that the dust in dominated by corundum and olivine (mean mass fractions of 46$\pm$6 and 31$\pm$11 \% respectively), followed by Mg-rich olivine and periclase (11$\pm$9 \% and 9$\pm$1 \%). The much lower abundance of Mg-rich olivine implies a non-negligible iron content in silicates. We find a mean crystallinity of 8$\pm$6 \% (range: 0--29\%) for our sample. While this is a small fraction of the total silicate dust mass, it is still higher than the crystallinity reported for the local interstellar medium (ISM) by \citet{Kemperetal2004} \citep[see also][]{LiDraine2001,LiZhaoLi2007}. We will discuss this result in the next section.\\

\section{Discussion}
\label{sec:discussion}
\begin{figure}
\includegraphics[width=\textwidth]{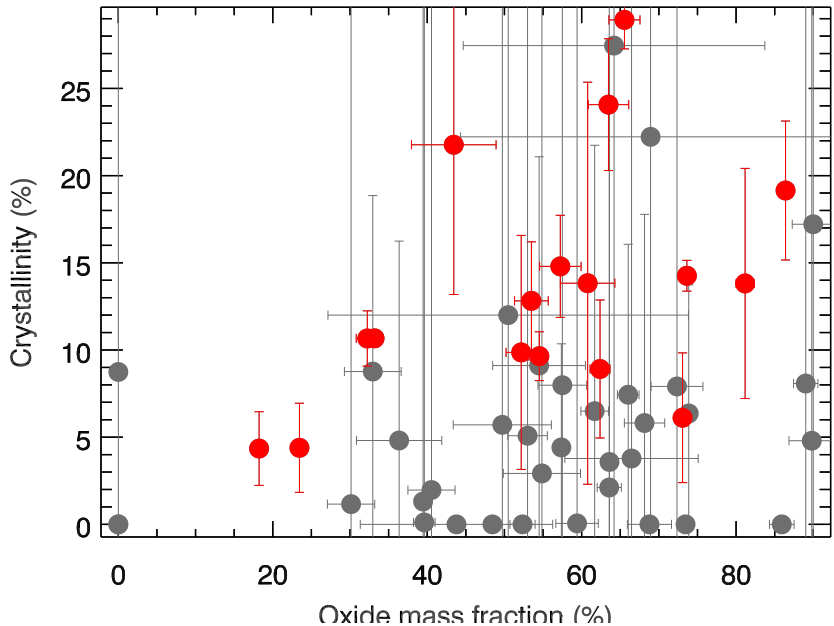}
\caption{Crystallinity plotted against the total mass fraction of oxides (corundum and periclase). The red symbols distinguish the objects for which the fractional uncertainty in the crystallinity is less than 100\%.
\label{fig:crysvsmfox}}
\end{figure}

The broader shape and redder peak position of the 9.7 $\mu$m silicate emission feature in AGN compared to Galactic sources has been attributed to a difference in dust properties -- either in shape, composition \citep{Siebenmorgenetal2005,Haoetal2005,Sturmetal2005}, or geometry of distribution \citep{Nenkova_02_Dust,Nikuttaetal2009}. Our finding that the mean crystallinity in the PG sample of quasars is higher than the upper limit derived for the local ISM supports this scenario. 
Our results are consistent with the classical dust condensation sequence. If the dust is distributed in clumps \citep[{\it e.g.},][]{PierKrolik1992,Nenkova_02_Dust}, the high abundance of alumina derived in this study might indicate that the dust condensation sequence is unable to proceed to the formation of large amounts of silicates, perhaps due to the clumps not being dense enough \citep[``freeze out"; {\it e.g.},][]{Tielens1990}. 
Moreover, Figure \ref{fig:crysvsmfox} shows that the crystallinity is in general higher for oxide-dominated dust (mass fraction $>$ 50\%, the median value for corundum; see Figure \ref{fig:boxplot}). Decreasing oxide fractions indicate advancing stages along the dust condensation sequence; the absence of pronounced crystallinity at very low oxide fractions might indicate the most advanced stage of the sequence, where a large amount of the dust has been processed into amorphous silicates.\\

\begin{figure}
\includegraphics[width=0.5\textwidth]{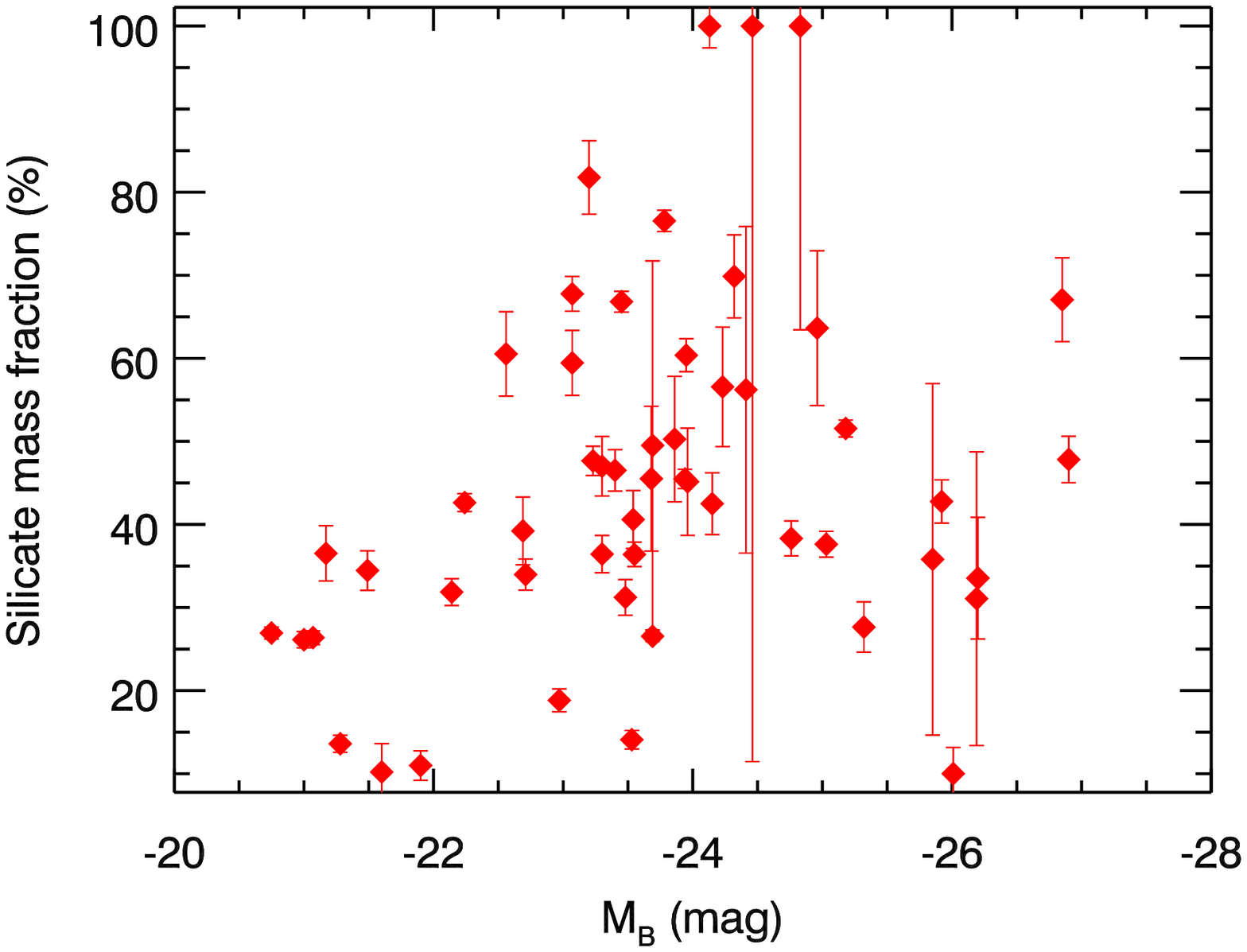}\includegraphics[width=0.5\textwidth]{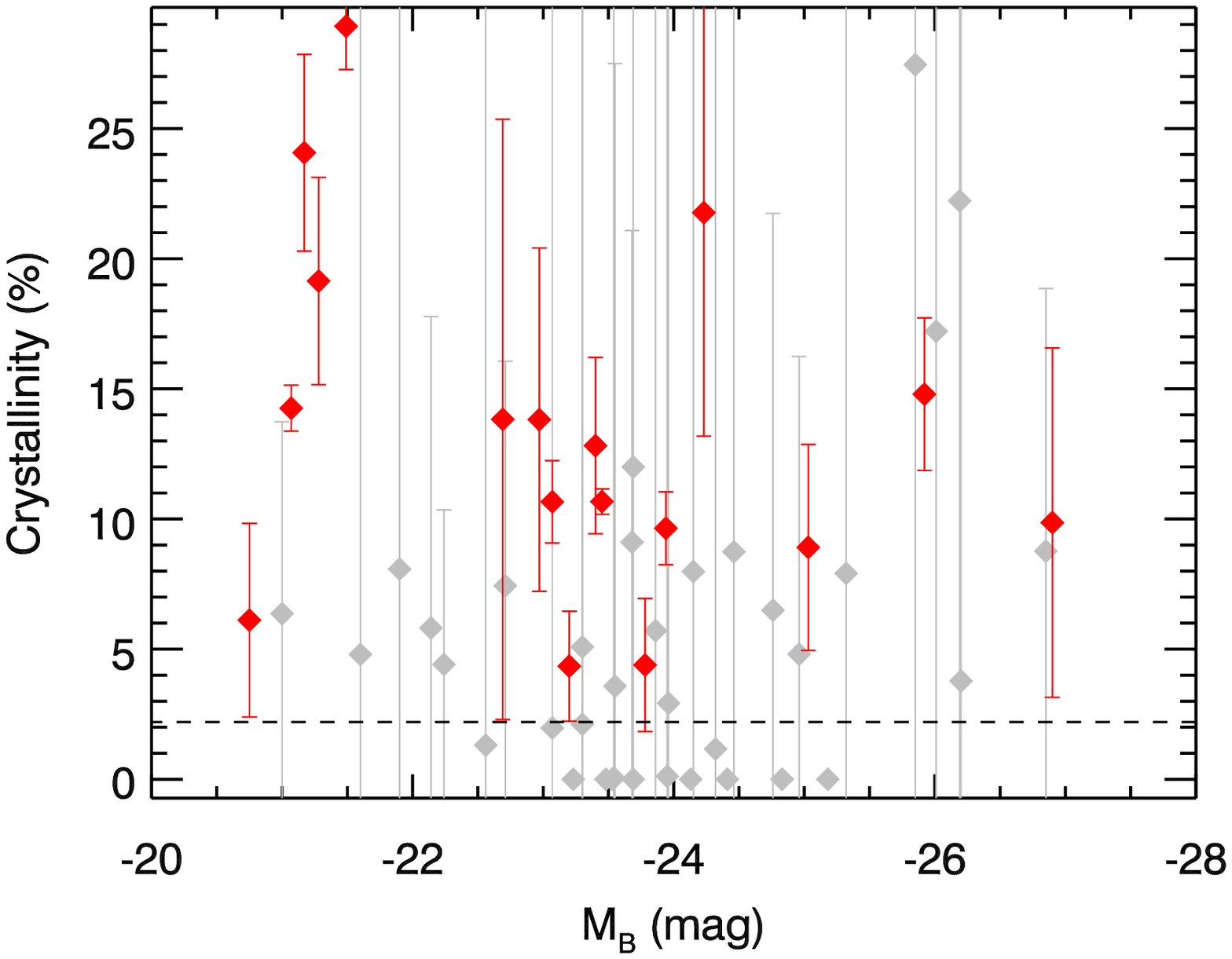}\\
\includegraphics[width=0.5\textwidth]{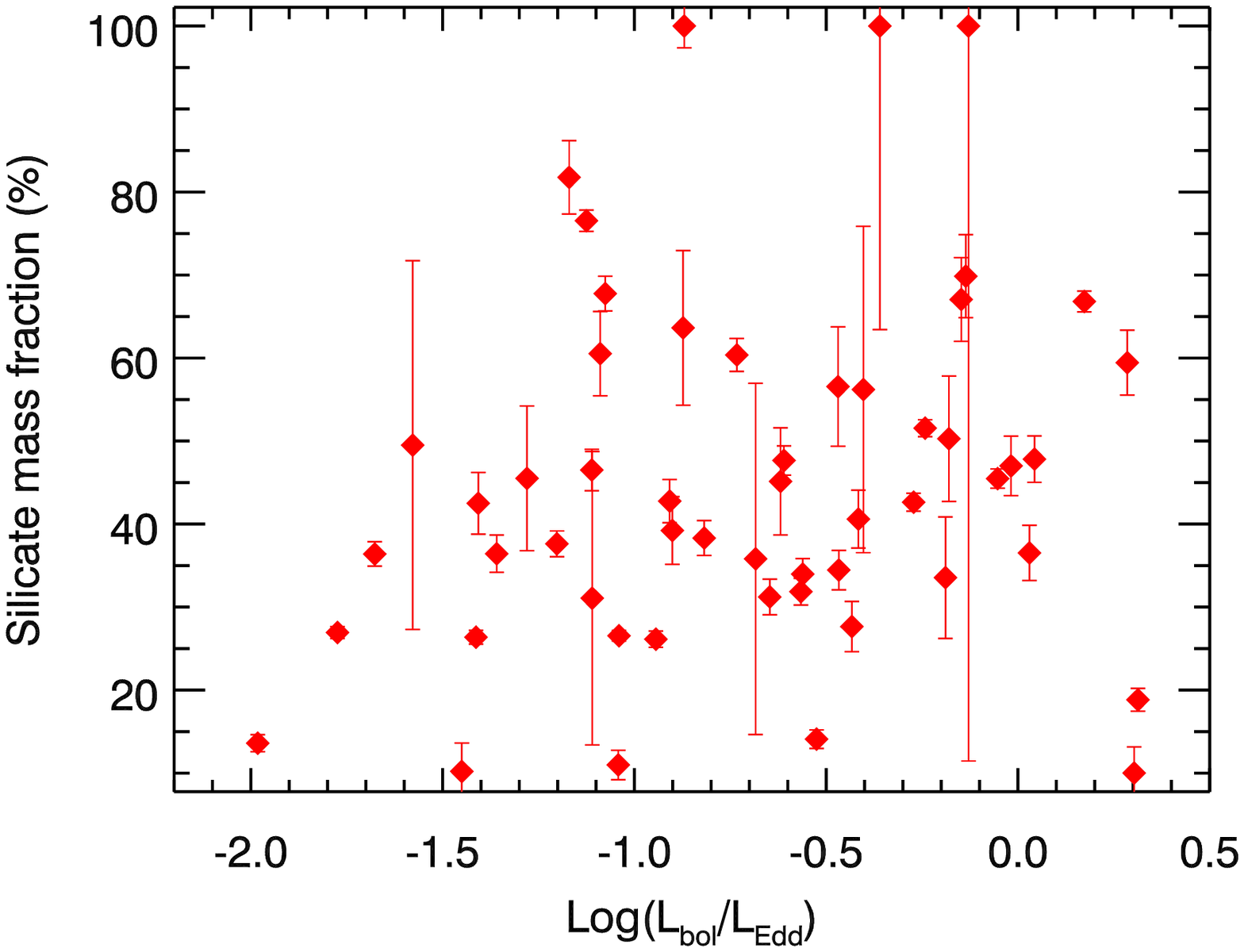}\includegraphics[width=0.5\textwidth]{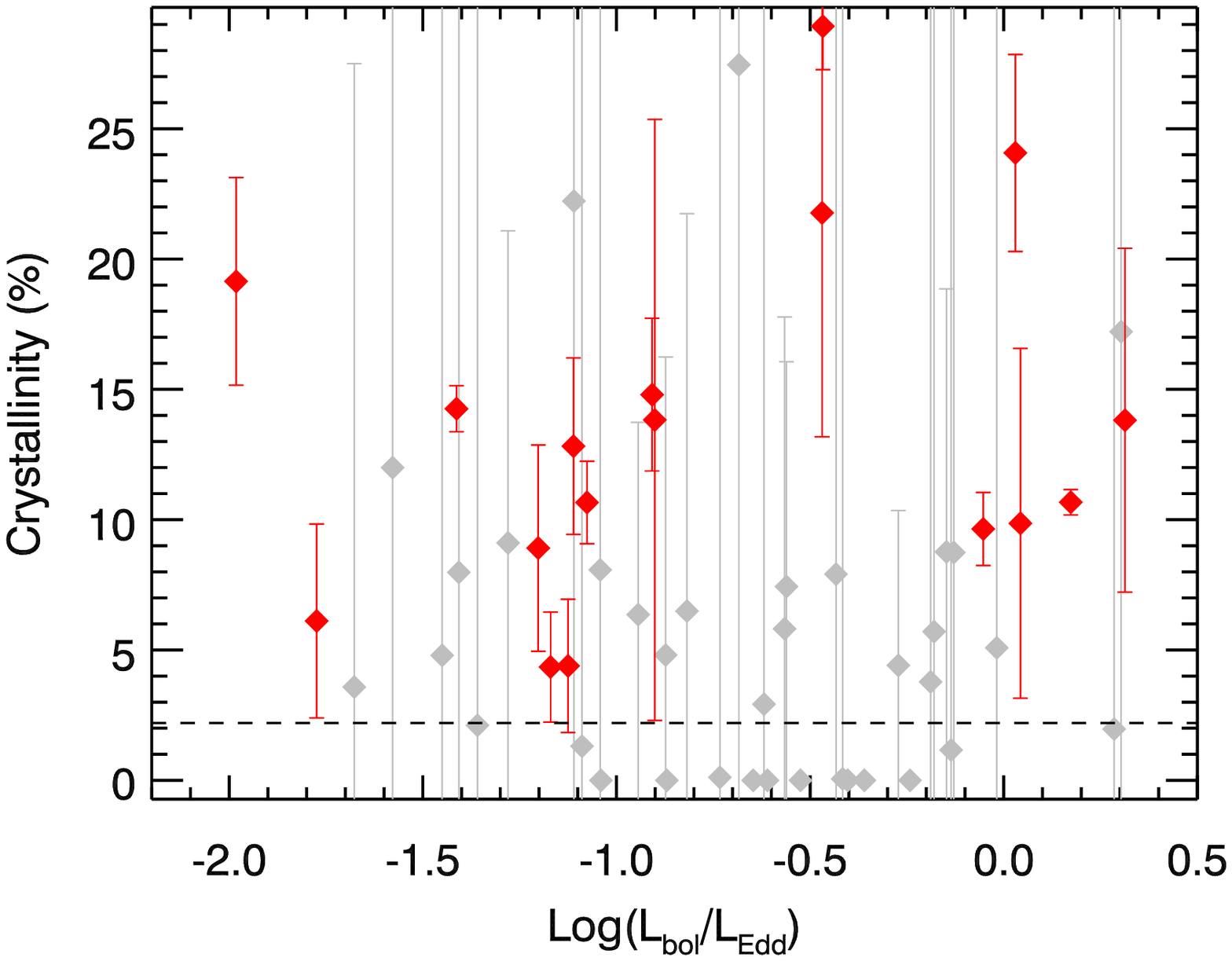}\\
\includegraphics[width=0.5\textwidth]{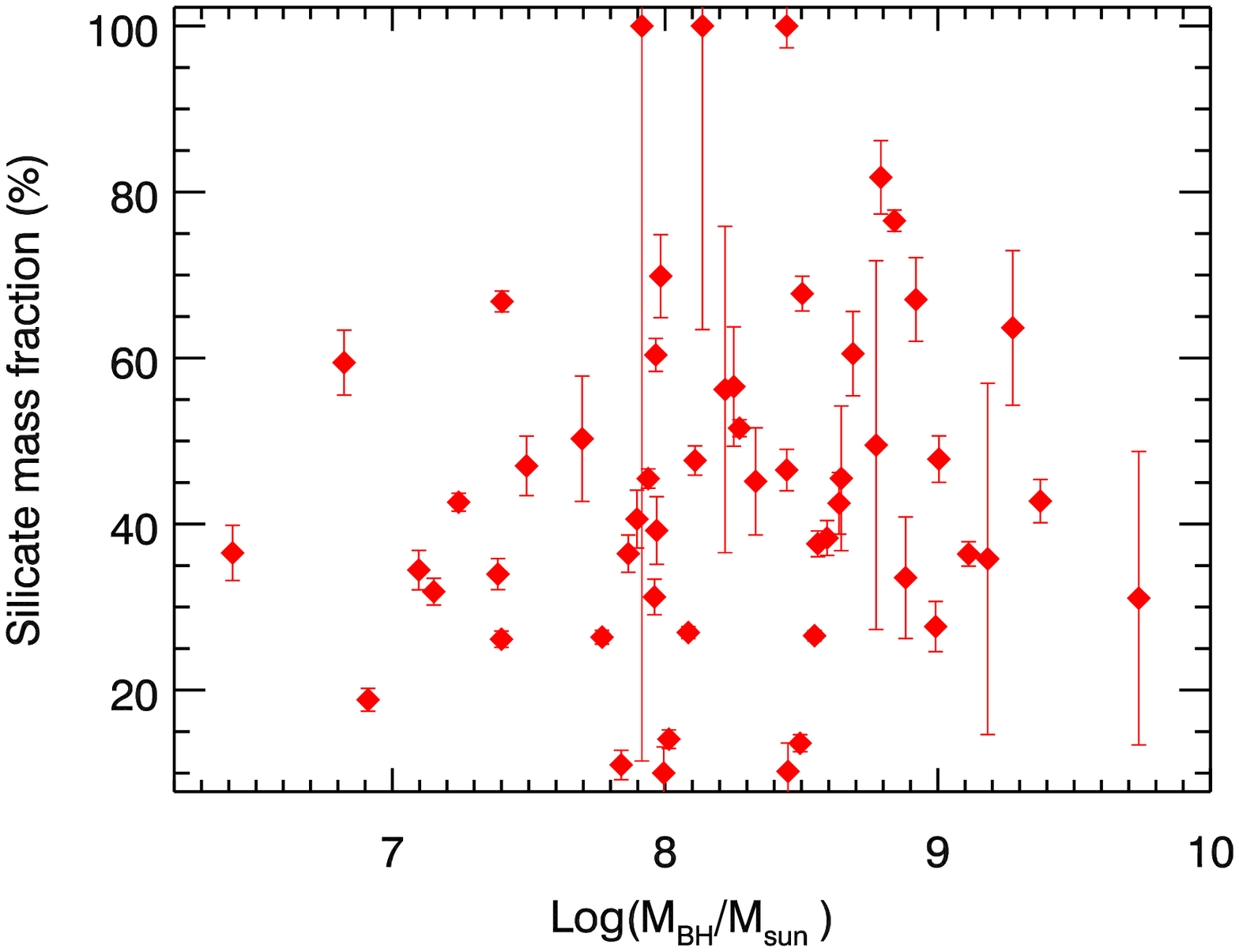}\includegraphics[width=0.5\textwidth]{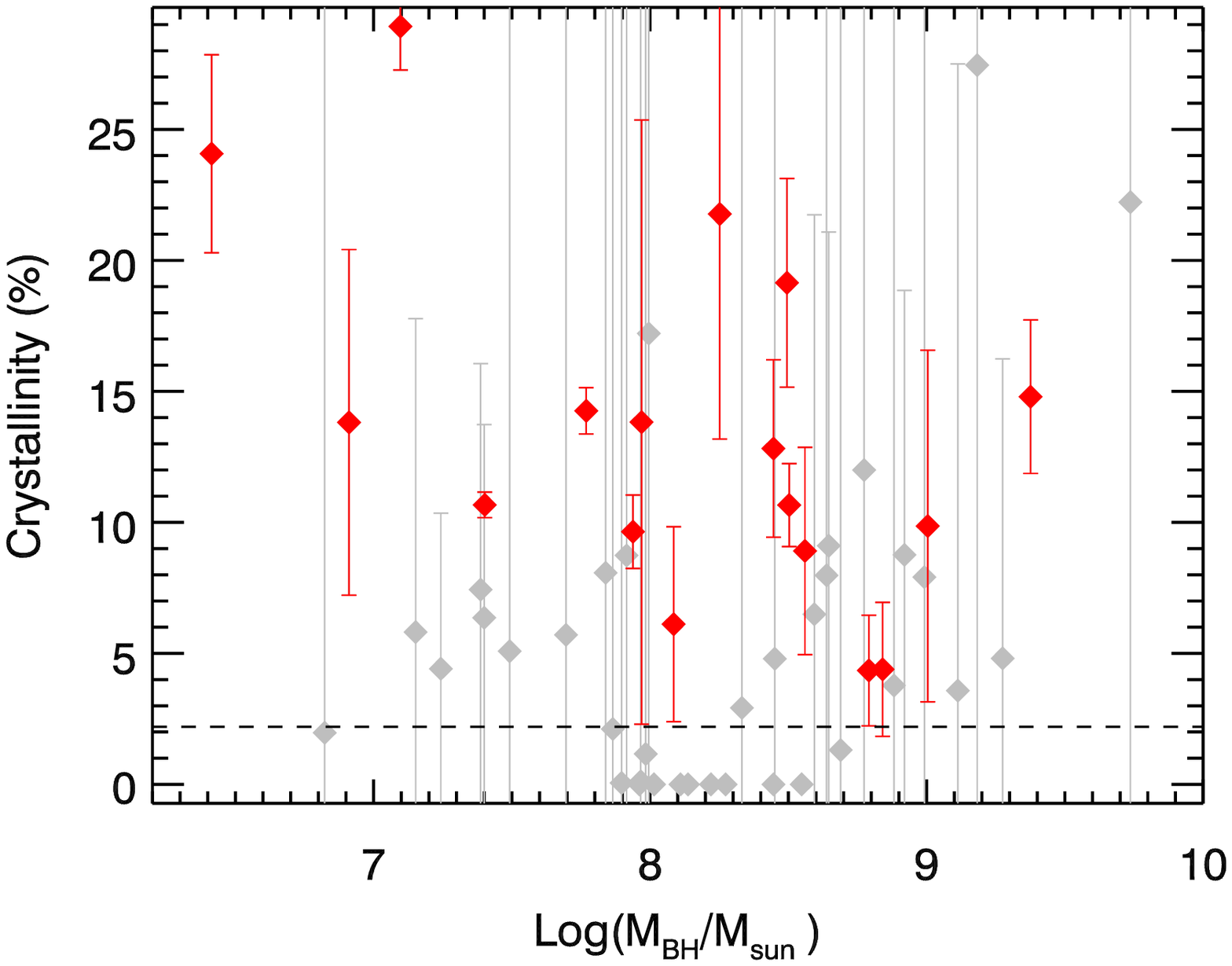}\\
\caption{{\em Left panels:} The total mass fraction of silicates (amorphous + crystalline) plotted versus the $B$-band absolute magnitude ({\em top}), the Eddington fraction ({\em middle}), and black hole mass ({\em bottom}). {\em Right panels:} The crystallinity versus the same parameters, with the good-quality fits (fractional uncertainty $<$ 100\%) shown in red. The horizontal dashed lines show the upper limit to the crystallinity derived for the local ISM by \citet{Kemperetal2004}.
\label{fig:fracvsphyspars}}
\end{figure}

We also investigate correlations between the physical parameters of the PG sample and the mass fractions derived in this study. Figure \ref{fig:fracvsphyspars} shows the total mass fraction of silicates, both amorphous and crystalline, versus the absolute $B$-band magnitude, the bolometric luminosity (in terms of the Eddington luminosity), and the black hole mass. The same plots are also shown for the crystallinity. Silicates dominate the dust composition for intermediate $B$-band luminosities ($-23~{\rm mag}>M_B>-25~{\rm mag}$), although this could also be attributed to the fact that there are more quasars in this luminosity range. A similar trend is observed with black hole mass. We do not see a clear trend of the crystallinity with luminosity, although the sources with higher good quality crystallinity determinations do occur at high luminosities. In general, higher luminosities correspond to enhanced densities associated with higher accretion rates, favoring the formation of crystalline silicates. A higher crystallinity could also be a result of dust processing in a wind, as radiation-driven winds are more important for luminous sources. If the dust is exposed directly to the quasar continuum for periods greater than the annealing timescale for amorphous grains \citep[$\sim$100 s for 0.1 $\mu$m silicate grains at 1600 K; see Fig. 1 in ][]{Gailetal1998}, it could favour the formation of crystalline species. The plots of crystallinity also compare the values computed in this paper with the local ISM determination of \citet{Kemperetal2004}. Crystallinities higher than the \citet{Kemperetal2004} limit occur over the entire range of luminosities and black hole masses. This is the first time that such a systematic result has been obtained in the analysis of dust around AGN.\\

While we have computed the crystallinity in 53 of the PG quasars, given the large number of parameters involved, and the fact that many of the spectra are noisy, we caution against overinterpreting any trends at this stage. It has been shown that the shape of the silicate features is influenced by properties such as grain size, shape and porosity (see, {\it e.g.}, Fig. 9 in \citealt{Dorschneretal1995}, Fig. 2 in \citealt{Minetal2005}, and \citealt{Smithetal2010}), and radiative transfer effects \citep[{\it e.g.},][]{Nikuttaetal2009}, which we have yet to explore. In subsequent papers, we will consider absorption efficiencies for a range of grain sizes, shapes, and mineralogies in order to refine our fits in order to take these aspects into account. The James Webb Space Telescope's improved sensitivity will provide higher quality spectra of these objects in the near future, allowing for more precise determinations of the dust composition around AGN.

\section{Conclusions}
\label{sec:conclusions}
We investigate the dust mineralogy of 86 quasars from the Palomar Green Sample with 10 $\mu$m silicate emission. By simultaneously fitting the infrared continuum as well as the dust features, we derive mass fractions of various species of amorphous and crystalline dust. After eliminating the noisiest spectra and bad fits from the sample, we are left with 53 quasars. For these, we find that most of the dust is in the form of alumina and amorphous silicates, with a small fraction in crystalline silicates. The crystallinity is higher when the dust is dominated by refractory oxides. The predominance of oxides may point to a freeze-out in the dust condensation sequence due to inadequate densities in the dust clumps. More than half of the quasars in our sample have crystallinities greater than 5\%, well above the upper limit for the local ISM. We find sources above this limit over the entire range of luminosities and black hole masses. This is the first time that such a result has been obtained for a large sample of quasars.\\

The results are sensitive to the various model assumptions; we are computing absorption efficiencies for a range of grain sizes, shapes, and mineralogies in order to refine these fits and take various grain properties and radiative transfer effects into account. Since the crystalline silicate features are narrow, they are somewhat less sensitive to the continuum determinations. The detection of crystallinity in our work is therefore robust. Future studies using JWST mid-infrared spectroscopy will potentially provide a systematic probe for the conditions in the dust condensation zone in quasar winds.\\

\section{Acknowledgements}
This research is supported by the Ministry of Science and Technology (MoST) of Taiwan, under grant number MOST104-2628-M-001-004-MY3. S.C.G. thanks the Natural Science and Engineering Research Council of Canada. This research has made use of the SIMBAD database, operated at CDS, Strasbourg, France.

\section*{References}
\bibliography{ms}

\begin{thebibliography}{67}
\expandafter\ifx\csname natexlab\endcsname\relax\def\natexlab#1{#1}\fi
\providecommand{\url}[1]{\texttt{#1}}
\providecommand{\href}[2]{#2}
\providecommand{\path}[1]{#1}
\providecommand{\DOIprefix}{doi:}
\providecommand{\ArXivprefix}{arXiv:}
\providecommand{\URLprefix}{URL: }
\providecommand{\Pubmedprefix}{pmid:}
\providecommand{\doi}[1]{\href{http://dx.doi.org/#1}{\path{#1}}}
\providecommand{\Pubmed}[1]{\href{pmid:#1}{\path{#1}}}
\providecommand{\bibinfo}[2]{#2}
\ifx\xfnm\relax \def\xfnm[#1]{\unskip,\space#1}\fi
\bibitem[{{Alonso-Herrero} et~al.(2016){Alonso-Herrero}, {Esquej}, {Roche},
  {Ramos Almeida}, {Gonz{\'a}lez-Mart{\'{\i}}n}, {Packham}, {Levenson},
  {Mason}, {Hern{\'a}n-Caballero}, {Pereira-Santaella}, {Alvarez}, {Aretxaga},
  {L{\'o}pez-Rodr{\'{\i}}guez}, {Colina}, {D{\'\i}az-Santos}, {Imanishi},
  {Rodr{\'\i}guez Espinosa} and {Perlman}}]{Alonso-Herrero_16_mid}
\bibinfo{author}{{Alonso-Herrero}, A.}, \bibinfo{author}{{Esquej}, P.},
  \bibinfo{author}{{Roche}, P.F.}, \bibinfo{author}{{Ramos Almeida}, C.},
  \bibinfo{author}{{Gonz{\'a}lez-Mart{\'{\i}}n}, O.},
  \bibinfo{author}{{Packham}, C.}, \bibinfo{author}{{Levenson}, N.A.},
  \bibinfo{author}{{Mason}, R.E.}, \bibinfo{author}{{Hern{\'a}n-Caballero},
  A.}, \bibinfo{author}{{Pereira-Santaella}, M.}, \bibinfo{author}{{Alvarez},
  C.}, \bibinfo{author}{{Aretxaga}, I.},
  \bibinfo{author}{{L{\'o}pez-Rodr{\'{\i}}guez}, E.},
  \bibinfo{author}{{Colina}, L.}, \bibinfo{author}{{D{\'\i}az-Santos}, T.},
  \bibinfo{author}{{Imanishi}, M.}, \bibinfo{author}{{Rodr{\'\i}guez Espinosa},
  J.M.}, \bibinfo{author}{{Perlman}, E.}, \bibinfo{year}{2016}.
\newblock \bibinfo{title}{A mid-infrared spectroscopic atlas of local active
  galactic nuclei on sub-arcsecond resolution using {GTC/CanariCam}}.
\newblock \bibinfo{journal}{\mnras} \bibinfo{volume}{455},
  \bibinfo{pages}{563--583}.
\newblock \DOIprefix\doi{10.1093/mnras/stv2342},
  \href{http://arxiv.org/abs/1510.02631}{\tt arXiv:1510.02631}.
\bibitem[{{Antonucci}(1993)}]{Antonucci_93_Unified}
\bibinfo{author}{{Antonucci}, R.}, \bibinfo{year}{1993}.
\newblock \bibinfo{title}{{Unified models for active galactic nuclei and
  quasars}}.
\newblock \bibinfo{journal}{\araa} \bibinfo{volume}{31},
  \bibinfo{pages}{473--521}.
\newblock \DOIprefix\doi{10.1146/annurev.aa.31.090193.002353}.
\bibitem[{{Baskin} and {Laor}(2005)}]{BaskinLaor2005}
\bibinfo{author}{{Baskin}, A.}, \bibinfo{author}{{Laor}, A.},
  \bibinfo{year}{2005}.
\newblock \bibinfo{title}{{What controls the CIV line profile in active
  galactic nuclei?}}
\newblock \bibinfo{journal}{\mnras} \bibinfo{volume}{356},
  \bibinfo{pages}{1029--1044}.
\newblock \DOIprefix\doi{10.1111/j.1365-2966.2004.08525.x},
  \href{http://arxiv.org/abs/astro-ph/0409196}{\tt arXiv:astro-ph/0409196}.
\bibitem[{{Begemann} et~al.(1997){Begemann}, {Dorschner}, {Henning},
  {Mutschke}, {G{\"u}rtler}, {K{\"o}mpe} and {Nass}}]{Begemannetal1997}
\bibinfo{author}{{Begemann}, B.}, \bibinfo{author}{{Dorschner}, J.},
  \bibinfo{author}{{Henning}, T.}, \bibinfo{author}{{Mutschke}, H.},
  \bibinfo{author}{{G{\"u}rtler}, J.}, \bibinfo{author}{{K{\"o}mpe}, C.},
  \bibinfo{author}{{Nass}, R.}, \bibinfo{year}{1997}.
\newblock \bibinfo{title}{{Aluminum Oxide and the Opacity of Oxygen-rich
  Circumstellar Dust in the 12-17 Micron Range}}.
\newblock \bibinfo{journal}{\apj} \bibinfo{volume}{476},
  \bibinfo{pages}{199--208}.
\bibitem[{Bianchi et~al.(2012)Bianchi, Maiolino and Risaliti}]{Bianchi_12_AGN}
\bibinfo{author}{Bianchi, S.}, \bibinfo{author}{Maiolino, R.},
  \bibinfo{author}{Risaliti, G.}, \bibinfo{year}{2012}.
\newblock \bibinfo{title}{{AGN Obscuration} and the {Unified Model}}.
\newblock \bibinfo{journal}{\aastr} \bibinfo{volume}{2012},
  \bibinfo{pages}{1--17}.
\newblock \URLprefix \url{http://www.hindawi.com/journals/aa/2012/782030/},
  \DOIprefix\doi{10.1155/2012/782030}.
\bibitem[{{Bohren} and {Huffman}(1983)}]{BohrenHuffman}
\bibinfo{author}{{Bohren}, C.F.}, \bibinfo{author}{{Huffman}, D.R.},
  \bibinfo{year}{1983}.
\newblock \bibinfo{title}{{Absorption and scattering of light by small
  particles}}.
\bibitem[{{Boroson} and {Green}(1992)}]{BorosonGreen1992}
\bibinfo{author}{{Boroson}, T.A.}, \bibinfo{author}{{Green}, R.F.},
  \bibinfo{year}{1992}.
\newblock \bibinfo{title}{{The emission-line properties of low-redshift
  quasi-stellar objects}}.
\newblock \bibinfo{journal}{\apjs} \bibinfo{volume}{80},
  \bibinfo{pages}{109--135}.
\newblock \DOIprefix\doi{10.1086/191661}.
\bibitem[{{Dorschner} et~al.(1995){Dorschner}, {Begemann}, {Henning}, {Jaeger}
  and {Mutschke}}]{Dorschneretal1995}
\bibinfo{author}{{Dorschner}, J.}, \bibinfo{author}{{Begemann}, B.},
  \bibinfo{author}{{Henning}, T.}, \bibinfo{author}{{Jaeger}, C.},
  \bibinfo{author}{{Mutschke}, H.}, \bibinfo{year}{1995}.
\newblock \bibinfo{title}{{Steps toward interstellar silicate mineralogy. II.
  Study of Mg-Fe-silicate glasses of variable composition.}}
\newblock \bibinfo{journal}{\aap} \bibinfo{volume}{300}, \bibinfo{pages}{503}.
\bibitem[{{Draine} and {Li}(2007)}]{Draine_07_Infrared}
\bibinfo{author}{{Draine}, B.T.}, \bibinfo{author}{{Li}, A.},
  \bibinfo{year}{2007}.
\newblock \bibinfo{title}{Infrared {Emission} from {Interstellar Dust}. {IV}.
  {The Silicate-Graphite-PAH Model} in the {Post-Spitzer Era}}.
\newblock \bibinfo{journal}{\apj} \bibinfo{volume}{657},
  \bibinfo{pages}{810--837}.
\newblock \DOIprefix\doi{10.1086/511055},
  \href{http://arxiv.org/abs/astro-ph/0608003}{\tt arXiv:astro-ph/0608003}.
\bibitem[{{Elitzur} and {Shlosman}(2006)}]{ElitzurShlosman2006}
\bibinfo{author}{{Elitzur}, M.}, \bibinfo{author}{{Shlosman}, I.},
  \bibinfo{year}{2006}.
\newblock \bibinfo{title}{{The AGN-obscuring Torus: The End of the ``Doughnut''
  Paradigm?}}
\newblock \bibinfo{journal}{\apjl} \bibinfo{volume}{648},
  \bibinfo{pages}{L101--L104}.
\newblock \DOIprefix\doi{10.1086/508158},
  \href{http://arxiv.org/abs/astro-ph/0605686}{\tt arXiv:astro-ph/0605686}.
\bibitem[{{Elvis} et~al.(2002){Elvis}, {Marengo} and
  {Karovska}}]{Elvisetal2002}
\bibinfo{author}{{Elvis}, M.}, \bibinfo{author}{{Marengo}, M.},
  \bibinfo{author}{{Karovska}, M.}, \bibinfo{year}{2002}.
\newblock \bibinfo{title}{{Smoking Quasars: A New Source for Cosmic Dust}}.
\newblock \bibinfo{journal}{\apjl} \bibinfo{volume}{567},
  \bibinfo{pages}{L107--L110}.
\newblock \DOIprefix\doi{10.1086/340006},
  \href{http://arxiv.org/abs/astro-ph/0202002}{\tt arXiv:astro-ph/0202002}.
\bibitem[{{Fritz} et~al.(2006){Fritz}, {Franceschini} and
  {Hatziminaoglou}}]{Fritz_06_Revisiting}
\bibinfo{author}{{Fritz}, J.}, \bibinfo{author}{{Franceschini}, A.},
  \bibinfo{author}{{Hatziminaoglou}, E.}, \bibinfo{year}{2006}.
\newblock \bibinfo{title}{Revisiting the infrared spectra of active galactic
  nuclei with a new torus emission model}.
\newblock \bibinfo{journal}{\mnras} \bibinfo{volume}{366},
  \bibinfo{pages}{767--786}.
\newblock \DOIprefix\doi{10.1111/j.1365-2966.2006.09866.x},
  \href{http://arxiv.org/abs/astro-ph/0511428}{\tt arXiv:astro-ph/0511428}.
\bibitem[{{Gail}(1998)}]{Gailetal1998}
\bibinfo{author}{{Gail}, H.P.}, \bibinfo{year}{1998}.
\newblock \bibinfo{title}{{Chemical reactions in protoplanetary accretion
  disks. IV. Multicomponent dust mixture}}.
\newblock \bibinfo{journal}{\aap} \bibinfo{volume}{332},
  \bibinfo{pages}{1099--1122}.
\bibitem[{{Gail} and {Sedlmayr}(1999)}]{GailSedlmayr1999}
\bibinfo{author}{{Gail}, H.P.}, \bibinfo{author}{{Sedlmayr}, E.},
  \bibinfo{year}{1999}.
\newblock \bibinfo{title}{{Mineral formation in stellar winds. I. Condensation
  sequence of silicate and iron grains in stationary oxygen rich outflows}}.
\newblock \bibinfo{journal}{\aap} \bibinfo{volume}{347},
  \bibinfo{pages}{594--616}.
\bibitem[{{Ge} et~al.(2016){Ge}, {Bian}, {Jiang}, {Liu} and
  {Wang}}]{Geetal2016}
\bibinfo{author}{{Ge}, X.}, \bibinfo{author}{{Bian}, W.H.},
  \bibinfo{author}{{Jiang}, X.L.}, \bibinfo{author}{{Liu}, W.S.},
  \bibinfo{author}{{Wang}, X.F.}, \bibinfo{year}{2016}.
\newblock \bibinfo{title}{{The underlying driver for the C IV Baldwin effect in
  QSOs with 0 $>$ z $>$ 5}}.
\newblock \bibinfo{journal}{\mnras} \bibinfo{volume}{462},
  \bibinfo{pages}{966--976}.
\newblock \DOIprefix\doi{10.1093/mnras/stw1605},
  \href{http://arxiv.org/abs/1608.02172}{\tt arXiv:1608.02172}.
\bibitem[{{Hao} et~al.(2005){Hao}, {Spoon}, {Sloan}, {Marshall}, {Armus},
  {Tielens}, {Sargent}, {van Bemmel}, {Charmandaris}, {Weedman} and
  {Houck}}]{Haoetal2005}
\bibinfo{author}{{Hao}, L.}, \bibinfo{author}{{Spoon}, H.W.W.},
  \bibinfo{author}{{Sloan}, G.C.}, \bibinfo{author}{{Marshall}, J.A.},
  \bibinfo{author}{{Armus}, L.}, \bibinfo{author}{{Tielens}, A.G.G.M.},
  \bibinfo{author}{{Sargent}, B.}, \bibinfo{author}{{van Bemmel}, I.M.},
  \bibinfo{author}{{Charmandaris}, V.}, \bibinfo{author}{{Weedman}, D.W.},
  \bibinfo{author}{{Houck}, J.R.}, \bibinfo{year}{2005}.
\newblock \bibinfo{title}{{The Detection of Silicate Emission from Quasars at
  10 and 18 Microns}}.
\newblock \bibinfo{journal}{\apjl} \bibinfo{volume}{625},
  \bibinfo{pages}{L75--L78}.
\newblock \DOIprefix\doi{10.1086/431227},
  \href{http://arxiv.org/abs/astro-ph/0504423}{\tt arXiv:astro-ph/0504423}.
\bibitem[{{Hao} et~al.(2007){Hao}, {Weedman}, {Spoon}, {Marshall}, {Levenson},
  {Elitzur} and {Houck}}]{Hao_07_Distributiona}
\bibinfo{author}{{Hao}, L.}, \bibinfo{author}{{Weedman}, D.W.},
  \bibinfo{author}{{Spoon}, H.W.W.}, \bibinfo{author}{{Marshall}, J.A.},
  \bibinfo{author}{{Levenson}, N.A.}, \bibinfo{author}{{Elitzur}, M.},
  \bibinfo{author}{{Houck}, J.R.}, \bibinfo{year}{2007}.
\newblock \bibinfo{title}{The distribution of silicate strength in spitzer
  spectra of {AGNs} and {ULIRGs}}.
\newblock \bibinfo{journal}{\apjl} \bibinfo{volume}{655},
  \bibinfo{pages}{L77--L80}.
\newblock \URLprefix \url{http://iopscience.iop.org/1538-4357/655/2/L77},
  \DOIprefix\doi{10.1086/511973},
  \href{http://arxiv.org/abs/astro-ph/0612509}{\tt arXiv:astro-ph/0612509}.
\bibitem[{{Hatziminaoglou} et~al.(2015){Hatziminaoglou},
  {Hern{\'a}n-Caballero}, {Feltre} and {Pi{\~n}ol
  Ferrer}}]{Hatziminaoglouetal2015}
\bibinfo{author}{{Hatziminaoglou}, E.},
  \bibinfo{author}{{Hern{\'a}n-Caballero}, A.}, \bibinfo{author}{{Feltre}, A.},
  \bibinfo{author}{{Pi{\~n}ol Ferrer}, N.}, \bibinfo{year}{2015}.
\newblock \bibinfo{title}{{A Complete Census of Silicate Features in the
  Mid-infrared Spectra of Active Galaxies}}.
\newblock \bibinfo{journal}{\apj} \bibinfo{volume}{803}, \bibinfo{pages}{110}.
\newblock \DOIprefix\doi{10.1088/0004-637X/803/2/110},
  \href{http://arxiv.org/abs/1502.05823}{\tt arXiv:1502.05823}.
\bibitem[{{Hofmeister} et~al.(2003){Hofmeister}, {Keppel} and
  {Speck}}]{Hofmeisteretal2003}
\bibinfo{author}{{Hofmeister}, A.M.}, \bibinfo{author}{{Keppel}, E.},
  \bibinfo{author}{{Speck}, A.K.}, \bibinfo{year}{2003}.
\newblock \bibinfo{title}{{Absorption and reflection infrared spectra of MgO
  and other diatomic compounds}}.
\newblock \bibinfo{journal}{\mnras} \bibinfo{volume}{345},
  \bibinfo{pages}{16--38}.
\newblock \DOIprefix\doi{10.1046/j.1365-8711.2003.06899.x}.
\bibitem[{{H{\"o}nig} et~al.(2006){H{\"o}nig}, {Beckert}, {Ohnaka} and
  {Weigelt}}]{Hoenig_06_Radiative}
\bibinfo{author}{{H{\"o}nig}, S.F.}, \bibinfo{author}{{Beckert}, T.},
  \bibinfo{author}{{Ohnaka}, K.}, \bibinfo{author}{{Weigelt}, G.},
  \bibinfo{year}{2006}.
\newblock \bibinfo{title}{{Radiative transfer modeling of three-dimensional
  clumpy AGN tori and its application to NGC 1068}}.
\newblock \bibinfo{journal}{\aap} \bibinfo{volume}{452},
  \bibinfo{pages}{459--471}.
\newblock \DOIprefix\doi{10.1051/0004-6361:20054622},
  \href{http://arxiv.org/abs/astro-ph/0602494}{\tt arXiv:astro-ph/0602494}.
\bibitem[{{Hony} et~al.(2001){Hony}, {Van Kerckhoven}, {Peeters}, {Tielens},
  {Hudgins} and {Allamandola}}]{Hony_01_CH}
\bibinfo{author}{{Hony}, S.}, \bibinfo{author}{{Van Kerckhoven}, C.},
  \bibinfo{author}{{Peeters}, E.}, \bibinfo{author}{{Tielens}, A.G.G.M.},
  \bibinfo{author}{{Hudgins}, D.M.}, \bibinfo{author}{{Allamandola}, L.J.},
  \bibinfo{year}{2001}.
\newblock \bibinfo{title}{{The CH out-of-plane bending modes of PAH molecules
  in astrophysical environments}}.
\newblock \bibinfo{journal}{\aap} \bibinfo{volume}{370},
  \bibinfo{pages}{1030--1043}.
\newblock \href{http://arxiv.org/abs/astro-ph/0103035}{\tt
  arXiv:astro-ph/0103035}.
\bibitem[{{Hopkins} et~al.(2012){Hopkins}, {Hayward}, {Narayanan} and
  {Hernquist}}]{Hopkins_12_origins}
\bibinfo{author}{{Hopkins}, P.F.}, \bibinfo{author}{{Hayward}, C.C.},
  \bibinfo{author}{{Narayanan}, D.}, \bibinfo{author}{{Hernquist}, L.},
  \bibinfo{year}{2012}.
\newblock \bibinfo{title}{The origins of active galactic nuclei obscuration:
  the 'torus' as a dynamical, unstable driver of accretion}.
\newblock \bibinfo{journal}{\mnras} \bibinfo{volume}{420},
  \bibinfo{pages}{320--339}.
\newblock \DOIprefix\doi{10.1111/j.1365-2966.2011.20035.x},
  \href{http://arxiv.org/abs/1108.3086}{\tt arXiv:1108.3086}.
\bibitem[{{Jaeger} et~al.(1998){Jaeger}, {Molster}, {Dorschner}, {Henning},
  {Mutschke} and {Waters}}]{Jaegeretal1998}
\bibinfo{author}{{Jaeger}, C.}, \bibinfo{author}{{Molster}, F.J.},
  \bibinfo{author}{{Dorschner}, J.}, \bibinfo{author}{{Henning}, T.},
  \bibinfo{author}{{Mutschke}, H.}, \bibinfo{author}{{Waters}, L.B.F.M.},
  \bibinfo{year}{1998}.
\newblock \bibinfo{title}{{Steps toward interstellar silicate mineralogy. IV.
  The crystalline revolution}}.
\newblock \bibinfo{journal}{\aap} \bibinfo{volume}{339},
  \bibinfo{pages}{904--916}.
\bibitem[{{Jaffe} et~al.(2004){Jaffe}, {Meisenheimer}, {R{\"o}ttgering},
  {Leinert}, {Richichi}, {Chesneau}, {Fraix-Burnet}, {Glazenborg-Kluttig},
  {Granato}, {Graser}, {Heijligers}, {K{\"o}hler}, {Malbet}, {Miley},
  {Paresce}, {Pel}, {Perrin}, {Przygodda}, {Schoeller}, {Sol}, {Waters},
  {Weigelt}, {Woillez} and {de Zeeuw}}]{Jaffe_04_central}
\bibinfo{author}{{Jaffe}, W.}, \bibinfo{author}{{Meisenheimer}, K.},
  \bibinfo{author}{{R{\"o}ttgering}, H.J.A.}, \bibinfo{author}{{Leinert}, C.},
  \bibinfo{author}{{Richichi}, A.}, \bibinfo{author}{{Chesneau}, O.},
  \bibinfo{author}{{Fraix-Burnet}, D.}, \bibinfo{author}{{Glazenborg-Kluttig},
  A.}, \bibinfo{author}{{Granato}, G.L.}, \bibinfo{author}{{Graser}, U.},
  \bibinfo{author}{{Heijligers}, B.}, \bibinfo{author}{{K{\"o}hler}, R.},
  \bibinfo{author}{{Malbet}, F.}, \bibinfo{author}{{Miley}, G.K.},
  \bibinfo{author}{{Paresce}, F.}, \bibinfo{author}{{Pel}, J.W.},
  \bibinfo{author}{{Perrin}, G.}, \bibinfo{author}{{Przygodda}, F.},
  \bibinfo{author}{{Schoeller}, M.}, \bibinfo{author}{{Sol}, H.},
  \bibinfo{author}{{Waters}, L.B.F.M.}, \bibinfo{author}{{Weigelt}, G.},
  \bibinfo{author}{{Woillez}, J.}, \bibinfo{author}{{de Zeeuw}, P.T.},
  \bibinfo{year}{2004}.
\newblock \bibinfo{title}{The central dusty torus in the active nucleus of
  {NGC} 1068}.
\newblock \bibinfo{journal}{\nat} \bibinfo{volume}{429},
  \bibinfo{pages}{47--49}.
\bibitem[{{J{\"a}ger} et~al.(2003){J{\"a}ger}, {Dorschner}, {Mutschke}, {Posch}
  and {Henning}}]{Jaegeretal2003}
\bibinfo{author}{{J{\"a}ger}, C.}, \bibinfo{author}{{Dorschner}, J.},
  \bibinfo{author}{{Mutschke}, H.}, \bibinfo{author}{{Posch}, T.},
  \bibinfo{author}{{Henning}, T.}, \bibinfo{year}{2003}.
\newblock \bibinfo{title}{{Steps toward interstellar silicate mineralogy. VII.
  Spectral properties and crystallization behaviour of magnesium silicates
  produced by the sol-gel method}}.
\newblock \bibinfo{journal}{\aap} \bibinfo{volume}{408},
  \bibinfo{pages}{193--204}.
\newblock \DOIprefix\doi{10.1051/0004-6361:20030916}.
\bibitem[{{Keating} et~al.(2012){Keating}, {Everett}, {Gallagher} and
  {Deo}}]{Keating+2012}
\bibinfo{author}{{Keating}, S.K.}, \bibinfo{author}{{Everett}, J.E.},
  \bibinfo{author}{{Gallagher}, S.C.}, \bibinfo{author}{{Deo}, R.P.},
  \bibinfo{year}{2012}.
\newblock \bibinfo{title}{{Sweeping Away the Mysteries of Dusty Continuous
  Winds in Active Galactic Nuclei}}.
\newblock \bibinfo{journal}{\apj} \bibinfo{volume}{749}, \bibinfo{pages}{32}.
\newblock \DOIprefix\doi{10.1088/0004-637X/749/1/32},
  \href{http://arxiv.org/abs/1202.4681}{\tt arXiv:1202.4681}.
\bibitem[{{Kemper} et~al.(2004){Kemper}, {Vriend} and
  {Tielens}}]{Kemperetal2004}
\bibinfo{author}{{Kemper}, F.}, \bibinfo{author}{{Vriend}, W.J.},
  \bibinfo{author}{{Tielens}, A.G.G.M.}, \bibinfo{year}{2004}.
\newblock \bibinfo{title}{{The Absence of Crystalline Silicates in the Diffuse
  Interstellar Medium}}.
\newblock \bibinfo{journal}{\apj} \bibinfo{volume}{609},
  \bibinfo{pages}{826--837}.
\newblock \DOIprefix\doi{10.1086/421339},
  \href{http://arxiv.org/abs/astro-ph/0403609}{\tt arXiv:astro-ph/0403609}.
\bibitem[{Klein and Dutrow(2008)}]{KleinDutrow2008}
\bibinfo{author}{Klein, C.}, \bibinfo{author}{Dutrow, B.},
  \bibinfo{year}{2008}.
\newblock \bibinfo{title}{Manual of Mineral Science, 23rd Edition}.
\newblock \bibinfo{publisher}{J. Wiley}, \bibinfo{address}{Hoboken, N.J}.
\bibitem[{{K{\"o}hler} and {Li}(2010)}]{KohlerLi2010}
\bibinfo{author}{{K{\"o}hler}, M.}, \bibinfo{author}{{Li}, A.},
  \bibinfo{year}{2010}.
\newblock \bibinfo{title}{{On the anomalous silicate absorption feature of the
  prototypical Seyfert 2 galaxy NGC1068}}.
\newblock \bibinfo{journal}{\mnras} \bibinfo{volume}{406},
  \bibinfo{pages}{L6--L10}.
\newblock \DOIprefix\doi{10.1111/j.1745-3933.2010.00870.x},
  \href{http://arxiv.org/abs/1210.6562}{\tt arXiv:1210.6562}.
\bibitem[{{K\"onigl} and {Kartje}(1994)}]{KoeniglKartje1994}
\bibinfo{author}{{K\"onigl}, A.}, \bibinfo{author}{{Kartje}, J.F.},
  \bibinfo{year}{1994}.
\newblock \bibinfo{title}{{Disk-driven hydromagnetic winds as a key ingredient
  of active galactic nuclei unification schemes}}.
\newblock \bibinfo{journal}{\apj} \bibinfo{volume}{434},
  \bibinfo{pages}{446--467}.
\newblock \DOIprefix\doi{10.1086/174746}.
\bibitem[{{Lebouteiller} et~al.(2015){Lebouteiller}, {Barry}, {Goes}, {Sloan},
  {Spoon}, {Weedman}, {Bernard-Salas} and {Houck}}]{Lebouteilleretal2015}
\bibinfo{author}{{Lebouteiller}, V.}, \bibinfo{author}{{Barry}, D.J.},
  \bibinfo{author}{{Goes}, C.}, \bibinfo{author}{{Sloan}, G.C.},
  \bibinfo{author}{{Spoon}, H.W.W.}, \bibinfo{author}{{Weedman}, D.W.},
  \bibinfo{author}{{Bernard-Salas}, J.}, \bibinfo{author}{{Houck}, J.R.},
  \bibinfo{year}{2015}.
\newblock \bibinfo{title}{{CASSIS: The Cornell Atlas of Spitzer/Infrared
  Spectrograph Sources. II. High-resolution Observations}}.
\newblock \bibinfo{journal}{\apjs} \bibinfo{volume}{218}, \bibinfo{pages}{21}.
\newblock \DOIprefix\doi{10.1088/0067-0049/218/2/21},
  \href{http://arxiv.org/abs/1506.07610}{\tt arXiv:1506.07610}.
\bibitem[{{Lebouteiller} et~al.(2011){Lebouteiller}, {Barry}, {Spoon},
  {Bernard-Salas}, {Sloan}, {Houck} and {Weedman}}]{Lebouteilleretal2011}
\bibinfo{author}{{Lebouteiller}, V.}, \bibinfo{author}{{Barry}, D.J.},
  \bibinfo{author}{{Spoon}, H.W.W.}, \bibinfo{author}{{Bernard-Salas}, J.},
  \bibinfo{author}{{Sloan}, G.C.}, \bibinfo{author}{{Houck}, J.R.},
  \bibinfo{author}{{Weedman}, D.W.}, \bibinfo{year}{2011}.
\newblock \bibinfo{title}{{CASSIS: The Cornell Atlas of Spitzer/Infrared
  Spectrograph Sources}}.
\newblock \bibinfo{journal}{\apjs} \bibinfo{volume}{196}, \bibinfo{pages}{8}.
\newblock \DOIprefix\doi{10.1088/0067-0049/196/1/8},
  \href{http://arxiv.org/abs/1108.3507}{\tt arXiv:1108.3507}.
\bibitem[{{Li} and {Draine}(2001)}]{LiDraine2001}
\bibinfo{author}{{Li}, A.}, \bibinfo{author}{{Draine}, B.T.},
  \bibinfo{year}{2001}.
\newblock \bibinfo{title}{{On Ultrasmall Silicate Grains in the Diffuse
  Interstellar Medium}}.
\newblock \bibinfo{journal}{\apjl} \bibinfo{volume}{550},
  \bibinfo{pages}{L213--L217}.
\newblock \DOIprefix\doi{10.1086/319640},
  \href{http://arxiv.org/abs/astro-ph/0012147}{\tt arXiv:astro-ph/0012147}.
\bibitem[{{Li} et~al.(2008){Li}, {Shi} and {Li}}]{Li_08_anomalous}
\bibinfo{author}{{Li}, M.P.}, \bibinfo{author}{{Shi}, Q.J.},
  \bibinfo{author}{{Li}, A.}, \bibinfo{year}{2008}.
\newblock \bibinfo{title}{On the anomalous silicate emission features of active
  galactic nuclei: a possible interpretation based on porous dust}.
\newblock \bibinfo{journal}{\mnras} \bibinfo{volume}{391},
  \bibinfo{pages}{L49--L53}.
\newblock \DOIprefix\doi{10.1111/j.1745-3933.2008.00553.x},
  \href{http://arxiv.org/abs/0808.4121}{\tt arXiv:0808.4121}.
\bibitem[{{Li} et~al.(2007){Li}, {Zhao} and {Li}}]{LiZhaoLi2007}
\bibinfo{author}{{Li}, M.P.}, \bibinfo{author}{{Zhao}, G.},
  \bibinfo{author}{{Li}, A.}, \bibinfo{year}{2007}.
\newblock \bibinfo{title}{{On the crystallinity of silicate dust in the
  interstellar medium}}.
\newblock \bibinfo{journal}{\mnras} \bibinfo{volume}{382},
  \bibinfo{pages}{L26--L29}.
\newblock \DOIprefix\doi{10.1111/j.1745-3933.2007.00382.x},
  \href{http://arxiv.org/abs/0808.4129}{\tt arXiv:0808.4129}.
\bibitem[{{Lodders} and {Fegley}(1999)}]{LoddersFegley1999}
\bibinfo{author}{{Lodders}, K.}, \bibinfo{author}{{Fegley}, Jr., B.},
  \bibinfo{year}{1999}.
\newblock \bibinfo{title}{{Condensation Chemistry of Circumstellar Grains}},
  in: \bibinfo{editor}{{Le Bertre}, T.}, \bibinfo{editor}{{Lebre}, A.},
  \bibinfo{editor}{{Waelkens}, C.} (Eds.), \bibinfo{booktitle}{Asymptotic Giant
  Branch Stars}, p. \bibinfo{pages}{279}.
\bibitem[{{L{\'o}pez-Gonzaga} et~al.(2014){L{\'o}pez-Gonzaga}, {Jaffe},
  {Burtscher}, {Tristram} and {Meisenheimer}}]{Lopez-Gonzaga_14_Revealing}
\bibinfo{author}{{L{\'o}pez-Gonzaga}, N.}, \bibinfo{author}{{Jaffe}, W.},
  \bibinfo{author}{{Burtscher}, L.}, \bibinfo{author}{{Tristram}, K.R.W.},
  \bibinfo{author}{{Meisenheimer}, K.}, \bibinfo{year}{2014}.
\newblock \bibinfo{title}{Revealing the large nuclear dust structures in {NGC}
  1068 with {MIDI/VLTI}}.
\newblock \bibinfo{journal}{\aap} \bibinfo{volume}{565}, \bibinfo{pages}{A71}.
\newblock \DOIprefix\doi{10.1051/0004-6361/201323002},
  \href{http://arxiv.org/abs/1401.3248}{\tt arXiv:1401.3248}.
\bibitem[{{Lopez-Rodriguez} et~al.(2016){Lopez-Rodriguez}, {Packham}, {Roche},
  {Alonso-Herrero}, {D{\'{\i}}az-Santos}, {Nikutta},
  {Gonz{\'a}lez-Mart{\'{\i}}n}, {{\'A}lvarez}, {Esquej}, {Espinosa}, {Perlman},
  {Ramos Almeida} and {Telesco}}]{Lopez-Rodriguez_16_Mid}
\bibinfo{author}{{Lopez-Rodriguez}, E.}, \bibinfo{author}{{Packham}, C.},
  \bibinfo{author}{{Roche}, P.F.}, \bibinfo{author}{{Alonso-Herrero}, A.},
  \bibinfo{author}{{D{\'{\i}}az-Santos}, T.}, \bibinfo{author}{{Nikutta}, R.},
  \bibinfo{author}{{Gonz{\'a}lez-Mart{\'{\i}}n}, O.},
  \bibinfo{author}{{{\'A}lvarez}, C.A.}, \bibinfo{author}{{Esquej}, P.},
  \bibinfo{author}{{Espinosa}, J.M.R.}, \bibinfo{author}{{Perlman}, E.},
  \bibinfo{author}{{Ramos Almeida}, C.}, \bibinfo{author}{{Telesco}, C.M.},
  \bibinfo{year}{2016}.
\newblock \bibinfo{title}{Mid-infrared imaging- and spectro-polarimetric
  subarcsecond observations of {NGC} 1068}.
\newblock \bibinfo{journal}{\mnras} \bibinfo{volume}{458},
  \bibinfo{pages}{3851--3866}.
\newblock \URLprefix \url{http://dx.doi.org/10.1093/mnras/stw541},
  \DOIprefix\doi{10.1093/mnras/stw541},
  \href{http://arxiv.org/abs/1603.01265}{\tt arXiv:1603.01265}.
\bibitem[{{Markwardt}(2009)}]{Markwardt2009}
\bibinfo{author}{{Markwardt}, C.B.}, \bibinfo{year}{2009}.
\newblock \bibinfo{title}{{Non-linear Least-squares Fitting in IDL with
  MPFIT}}, in: \bibinfo{editor}{{Bohlender}, D.A.}, \bibinfo{editor}{{Durand},
  D.}, \bibinfo{editor}{{Dowler}, P.} (Eds.), \bibinfo{booktitle}{Astronomical
  Data Analysis Software and Systems XVIII}, p. \bibinfo{pages}{251}.
\newblock \href{http://arxiv.org/abs/0902.2850}{\tt arXiv:0902.2850}.
\bibitem[{{Markwick-Kemper} et~al.(2007){Markwick-Kemper}, {Gallagher}, {Hines}
  and {Bouwman}}]{Markwick-Kemperetal2007}
\bibinfo{author}{{Markwick-Kemper}, F.}, \bibinfo{author}{{Gallagher}, S.C.},
  \bibinfo{author}{{Hines}, D.C.}, \bibinfo{author}{{Bouwman}, J.},
  \bibinfo{year}{2007}.
\newblock \bibinfo{title}{{Dust in the Wind: Crystalline Silicates, Corundum,
  and Periclase in PG 2112+059}}.
\newblock \bibinfo{journal}{\apjl} \bibinfo{volume}{668},
  \bibinfo{pages}{L107--L110}.
\newblock \DOIprefix\doi{10.1086/523104},
  \href{http://arxiv.org/abs/0710.2225}{\tt arXiv:0710.2225}.
\bibitem[{{Mason}(2015)}]{Mason_15_Dust}
\bibinfo{author}{{Mason}, R.E.}, \bibinfo{year}{2015}.
\newblock \bibinfo{title}{Dust in the torus of the {AGN} unified model}.
\newblock \bibinfo{journal}{\planss} \bibinfo{volume}{116},
  \bibinfo{pages}{97--101}.
\newblock \URLprefix
  \url{http://linkinghub.elsevier.com/retrieve/pii/S0032063315000483},
  \DOIprefix\doi{10.1016/j.pss.2015.02.013},
  \href{http://arxiv.org/abs/1412.6189}{\tt arXiv:1412.6189}.
\bibitem[{{Mason} et~al.(2006){Mason}, {Geballe}, {Packham}, {Levenson},
  {Elitzur}, {Fisher} and {Perlman}}]{Mason_06_Spatially}
\bibinfo{author}{{Mason}, R.E.}, \bibinfo{author}{{Geballe}, T.R.},
  \bibinfo{author}{{Packham}, C.}, \bibinfo{author}{{Levenson}, N.A.},
  \bibinfo{author}{{Elitzur}, M.}, \bibinfo{author}{{Fisher}, R.S.},
  \bibinfo{author}{{Perlman}, E.}, \bibinfo{year}{2006}.
\newblock \bibinfo{title}{Spatially resolved {Mid-Infrared} spectroscopy of
  {NGC} 1068: The nature and distribution of the nuclear material}.
\newblock \bibinfo{journal}{\apj} \bibinfo{volume}{640},
  \bibinfo{pages}{612--624}.
\newblock \DOIprefix\doi{10.1086/500299},
  \href{http://arxiv.org/abs/astro-ph/0512202}{\tt arXiv:astro-ph/0512202}.
\bibitem[{{Min} et~al.(2005){Min}, {Hovenier} and {de Koter}}]{Minetal2005}
\bibinfo{author}{{Min}, M.}, \bibinfo{author}{{Hovenier}, J.W.},
  \bibinfo{author}{{de Koter}, A.}, \bibinfo{year}{2005}.
\newblock \bibinfo{title}{{Modeling optical properties of cosmic dust grains
  using a distribution of hollow spheres}}.
\newblock \bibinfo{journal}{\aap} \bibinfo{volume}{432},
  \bibinfo{pages}{909--920}.
\newblock \DOIprefix\doi{10.1051/0004-6361:20041920},
  \href{http://arxiv.org/abs/astro-ph/0503068}{\tt arXiv:astro-ph/0503068}.
\bibitem[{{Nenkova} et~al.(2002){Nenkova}, {Ivezi{\'c}} and
  {Elitzur}}]{Nenkova_02_Dust}
\bibinfo{author}{{Nenkova}, M.}, \bibinfo{author}{{Ivezi{\'c}}, {\v Z}.},
  \bibinfo{author}{{Elitzur}, M.}, \bibinfo{year}{2002}.
\newblock \bibinfo{title}{Dust {Emission} from {Active Galactic Nuclei}}.
\newblock \bibinfo{journal}{\apjl} \bibinfo{volume}{570},
  \bibinfo{pages}{L9--L12}.
\newblock \URLprefix \url{http://iopscience.iop.org/1538-4357/570/1/L9},
  \DOIprefix\doi{10.1086/340857},
  \href{http://arxiv.org/abs/astro-ph/0202405}{\tt arXiv:astro-ph/0202405}.
\bibitem[{{Netzer}(2015)}]{Netzer_15_Revisiting}
\bibinfo{author}{{Netzer}, H.}, \bibinfo{year}{2015}.
\newblock \bibinfo{title}{Revisiting the {Unified Model} of {Active Galactic
  Nuclei}}.
\newblock \bibinfo{journal}{\araa} \bibinfo{volume}{53},
  \bibinfo{pages}{365--408}.
\newblock \DOIprefix\doi{10.1146/annurev-astro-082214-122302},
  \href{http://arxiv.org/abs/1505.00811}{\tt arXiv:1505.00811}.
\bibitem[{{Netzer} et~al.(2007){Netzer}, {Lutz}, {Schweitzer}, {Contursi},
  {Sturm}, {Tacconi}, {Veilleux}, {Kim}, {Rupke}, {Baker}, {Dasyra},
  {Mazzarella} and {Lord}}]{Netzeretal2007}
\bibinfo{author}{{Netzer}, H.}, \bibinfo{author}{{Lutz}, D.},
  \bibinfo{author}{{Schweitzer}, M.}, \bibinfo{author}{{Contursi}, A.},
  \bibinfo{author}{{Sturm}, E.}, \bibinfo{author}{{Tacconi}, L.J.},
  \bibinfo{author}{{Veilleux}, S.}, \bibinfo{author}{{Kim}, D.C.},
  \bibinfo{author}{{Rupke}, D.}, \bibinfo{author}{{Baker}, A.J.},
  \bibinfo{author}{{Dasyra}, K.}, \bibinfo{author}{{Mazzarella}, J.},
  \bibinfo{author}{{Lord}, S.}, \bibinfo{year}{2007}.
\newblock \bibinfo{title}{{Spitzer Quasar and ULIRG Evolution Study (QUEST).
  II. The Spectral Energy Distributions of Palomar-Green Quasars}}.
\newblock \bibinfo{journal}{\apj} \bibinfo{volume}{666},
  \bibinfo{pages}{806--816}.
\newblock \DOIprefix\doi{10.1086/520716},
  \href{http://arxiv.org/abs/0706.0818}{\tt arXiv:0706.0818}.
\bibitem[{{Nikutta} et~al.(2009){Nikutta}, {Elitzur} and
  {Lacy}}]{Nikuttaetal2009}
\bibinfo{author}{{Nikutta}, R.}, \bibinfo{author}{{Elitzur}, M.},
  \bibinfo{author}{{Lacy}, M.}, \bibinfo{year}{2009}.
\newblock \bibinfo{title}{{On the 10 {$\mu$}m Silicate Feature in Active
  Galactic Nuclei}}.
\newblock \bibinfo{journal}{\apj} \bibinfo{volume}{707},
  \bibinfo{pages}{1550--1559}.
\newblock \DOIprefix\doi{10.1088/0004-637X/707/2/1550},
  \href{http://arxiv.org/abs/0910.5521}{\tt arXiv:0910.5521}.
\bibitem[{{Petric} et~al.(2015){Petric}, {Ho}, {Flagey} and
  {Scoville}}]{Petricetal2015}
\bibinfo{author}{{Petric}, A.O.}, \bibinfo{author}{{Ho}, L.C.},
  \bibinfo{author}{{Flagey}, N.J.M.}, \bibinfo{author}{{Scoville}, N.Z.},
  \bibinfo{year}{2015}.
\newblock \bibinfo{title}{{Herschel Survey of the Palomar-Green QSOs at Low
  Redshift}}.
\newblock \bibinfo{journal}{\apjs} \bibinfo{volume}{219}, \bibinfo{pages}{22}.
\newblock \DOIprefix\doi{10.1088/0067-0049/219/2/22},
  \href{http://arxiv.org/abs/1505.05273}{\tt arXiv:1505.05273}.
\bibitem[{{Pier} and {Krolik}(1992)}]{PierKrolik1992}
\bibinfo{author}{{Pier}, E.A.}, \bibinfo{author}{{Krolik}, J.H.},
  \bibinfo{year}{1992}.
\newblock \bibinfo{title}{{Infrared spectra of obscuring dust tori around
  active galactic nuclei. I - Calculational method and basic trends}}.
\newblock \bibinfo{journal}{\apj} \bibinfo{volume}{401},
  \bibinfo{pages}{99--109}.
\newblock \DOIprefix\doi{10.1086/172042}.
\bibitem[{{Poncelet} et~al.(2006){Poncelet}, {Perrin} and
  {Sol}}]{Poncelet_06_new}
\bibinfo{author}{{Poncelet}, A.}, \bibinfo{author}{{Perrin}, G.},
  \bibinfo{author}{{Sol}, H.}, \bibinfo{year}{2006}.
\newblock \bibinfo{title}{A new analysis of the nucleus of {NGC} 1068 with
  {MIDI} observations}.
\newblock \bibinfo{journal}{\aap} \bibinfo{volume}{450},
  \bibinfo{pages}{483--494}.
\newblock \DOIprefix\doi{10.1051/0004-6361:20053608},
  \href{http://arxiv.org/abs/astro-ph/0512560}{\tt arXiv:astro-ph/0512560}.
\bibitem[{{Rhee} and {Larkin}(2006)}]{Rhee_06_First}
\bibinfo{author}{{Rhee}, J.H.}, \bibinfo{author}{{Larkin}, J.E.},
  \bibinfo{year}{2006}.
\newblock \bibinfo{title}{The {First Spatially Resolved Mid-Infrared Spectra}
  of {NGC} 1068 {Obtained} at {Diffraction}-limited {Resolution} with the {Keck
  I Telescope Long Wavelength Spectrometer}}.
\newblock \bibinfo{journal}{\apj} \bibinfo{volume}{640},
  \bibinfo{pages}{625--638}.
\newblock \DOIprefix\doi{10.1086/500122},
  \href{http://arxiv.org/abs/astro-ph/0512050}{\tt arXiv:astro-ph/0512050}.
\bibitem[{{Sales} et~al.(2011){Sales}, {Pastoriza}, {Riffel}, {Winge},
  {Rodr{\'{\i}}guez-Ardila} and {Carciofi}}]{Sales_11_Compton}
\bibinfo{author}{{Sales}, D.A.}, \bibinfo{author}{{Pastoriza}, M.G.},
  \bibinfo{author}{{Riffel}, R.}, \bibinfo{author}{{Winge}, C.},
  \bibinfo{author}{{Rodr{\'{\i}}guez-Ardila}, A.}, \bibinfo{author}{{Carciofi},
  A.C.}, \bibinfo{year}{2011}.
\newblock \bibinfo{title}{{The Compton-thick Seyfert 2~{N}ucleus of NGC 3281:
  Torus Constraints from the 9.7 {$\mu$}m Silicate Absorption}}.
\newblock \bibinfo{journal}{\apj} \bibinfo{volume}{738}, \bibinfo{pages}{109}.
\newblock \URLprefix \url{http://iopscience.iop.org/0004-637X/738/1/109},
  \DOIprefix\doi{10.1088/0004-637X/738/1/109},
  \href{http://arxiv.org/abs/1106.5731}{\tt arXiv:1106.5731}.
\bibitem[{{Schmidt} and {Green}(1983)}]{SchmidtGreen1983}
\bibinfo{author}{{Schmidt}, M.}, \bibinfo{author}{{Green}, R.F.},
  \bibinfo{year}{1983}.
\newblock \bibinfo{title}{{Quasar evolution derived from the Palomar bright
  quasar survey and other complete quasar surveys}}.
\newblock \bibinfo{journal}{\apj} \bibinfo{volume}{269},
  \bibinfo{pages}{352--374}.
\newblock \DOIprefix\doi{10.1086/161048}.
\bibitem[{{Schweitzer} et~al.(2006){Schweitzer}, {Lutz}, {Sturm}, {Contursi},
  {Tacconi}, {Lehnert}, {Dasyra}, {Genzel}, {Veilleux}, {Rupke}, {Kim},
  {Baker}, {Netzer}, {Sternberg}, {Mazzarella} and {Lord}}]{Schweitzeretal2006}
\bibinfo{author}{{Schweitzer}, M.}, \bibinfo{author}{{Lutz}, D.},
  \bibinfo{author}{{Sturm}, E.}, \bibinfo{author}{{Contursi}, A.},
  \bibinfo{author}{{Tacconi}, L.J.}, \bibinfo{author}{{Lehnert}, M.D.},
  \bibinfo{author}{{Dasyra}, K.M.}, \bibinfo{author}{{Genzel}, R.},
  \bibinfo{author}{{Veilleux}, S.}, \bibinfo{author}{{Rupke}, D.},
  \bibinfo{author}{{Kim}, D.C.}, \bibinfo{author}{{Baker}, A.J.},
  \bibinfo{author}{{Netzer}, H.}, \bibinfo{author}{{Sternberg}, A.},
  \bibinfo{author}{{Mazzarella}, J.}, \bibinfo{author}{{Lord}, S.},
  \bibinfo{year}{2006}.
\newblock \bibinfo{title}{{Spitzer Quasar and ULIRG Evolution Study (QUEST). I.
  The Origin of the Far-Infrared Continuum of QSOs}}.
\newblock \bibinfo{journal}{\apj} \bibinfo{volume}{649},
  \bibinfo{pages}{79--90}.
\newblock \DOIprefix\doi{10.1086/506510},
  \href{http://arxiv.org/abs/astro-ph/0606158}{\tt arXiv:astro-ph/0606158}.
\bibitem[{{Shi} et~al.(2006){Shi}, {Rieke}, {Hines}, {Gorjian}, {Werner},
  {Cleary}, {Low}, {Smith} and {Bouwman}}]{Shietal2006}
\bibinfo{author}{{Shi}, Y.}, \bibinfo{author}{{Rieke}, G.H.},
  \bibinfo{author}{{Hines}, D.C.}, \bibinfo{author}{{Gorjian}, V.},
  \bibinfo{author}{{Werner}, M.W.}, \bibinfo{author}{{Cleary}, K.},
  \bibinfo{author}{{Low}, F.J.}, \bibinfo{author}{{Smith}, P.S.},
  \bibinfo{author}{{Bouwman}, J.}, \bibinfo{year}{2006}.
\newblock \bibinfo{title}{{9.7 {$\mu$}m Silicate Features in Active Galactic
  Nuclei: New Insights into Unification Models}}.
\newblock \bibinfo{journal}{\apj} \bibinfo{volume}{653},
  \bibinfo{pages}{127--136}.
\newblock \DOIprefix\doi{10.1086/508737},
  \href{http://arxiv.org/abs/astro-ph/0608645}{\tt arXiv:astro-ph/0608645}.
\bibitem[{{Shi} et~al.(2014){Shi}, {Rieke}, {Ogle}, {Su} and
  {Balog}}]{Shietal2014}
\bibinfo{author}{{Shi}, Y.}, \bibinfo{author}{{Rieke}, G.H.},
  \bibinfo{author}{{Ogle}, P.M.}, \bibinfo{author}{{Su}, K.Y.L.},
  \bibinfo{author}{{Balog}, Z.}, \bibinfo{year}{2014}.
\newblock \bibinfo{title}{{Infrared Spectra and Photometry Of Complete Samples
  of Palomar-Green and Two Micron All Sky Survey Quasars}}.
\newblock \bibinfo{journal}{\apjs} \bibinfo{volume}{214}, \bibinfo{pages}{23}.
\newblock \DOIprefix\doi{10.1088/0067-0049/214/2/23},
  \href{http://arxiv.org/abs/1408.5909}{\tt arXiv:1408.5909}.
\bibitem[{{Siebenmorgen} et~al.(2005){Siebenmorgen}, {Haas}, {Kr{\"u}gel} and
  {Schulz}}]{Siebenmorgenetal2005}
\bibinfo{author}{{Siebenmorgen}, R.}, \bibinfo{author}{{Haas}, M.},
  \bibinfo{author}{{Kr{\"u}gel}, E.}, \bibinfo{author}{{Schulz}, B.},
  \bibinfo{year}{2005}.
\newblock \bibinfo{title}{{Discovery of 10 {$\mu$}m silicate emission in
  quasars. Evidence of the AGN unification scheme}}.
\newblock \bibinfo{journal}{\aap} \bibinfo{volume}{436},
  \bibinfo{pages}{L5--L8}.
\newblock \DOIprefix\doi{10.1051/0004-6361:200500109}.
\bibitem[{{Smith} et~al.(2010){Smith}, {Li}, {Li}, {K{\"o}hler}, {Ashby},
  {Fazio}, {Huang}, {Marengo}, {Wang}, {Willner}, {Zezas}, {Spinoglio} and
  {Wu}}]{Smithetal2010}
\bibinfo{author}{{Smith}, H.A.}, \bibinfo{author}{{Li}, A.},
  \bibinfo{author}{{Li}, M.P.}, \bibinfo{author}{{K{\"o}hler}, M.},
  \bibinfo{author}{{Ashby}, M.L.N.}, \bibinfo{author}{{Fazio}, G.G.},
  \bibinfo{author}{{Huang}, J.S.}, \bibinfo{author}{{Marengo}, M.},
  \bibinfo{author}{{Wang}, Z.}, \bibinfo{author}{{Willner}, S.},
  \bibinfo{author}{{Zezas}, A.}, \bibinfo{author}{{Spinoglio}, L.},
  \bibinfo{author}{{Wu}, Y.L.}, \bibinfo{year}{2010}.
\newblock \bibinfo{title}{{Anomalous Silicate Dust Emission in the Type 1 Liner
  Nucleus of M81}}.
\newblock \bibinfo{journal}{\apj} \bibinfo{volume}{716},
  \bibinfo{pages}{490--503}.
\newblock \DOIprefix\doi{10.1088/0004-637X/716/1/490},
  \href{http://arxiv.org/abs/1004.2277}{\tt arXiv:1004.2277}.
\bibitem[{{Stalevski} et~al.(2012){Stalevski}, {Fritz}, {Baes}, {Nakos} and
  {Popovi{\'c}}}]{Stalevski_12_3D}
\bibinfo{author}{{Stalevski}, M.}, \bibinfo{author}{{Fritz}, J.},
  \bibinfo{author}{{Baes}, M.}, \bibinfo{author}{{Nakos}, T.},
  \bibinfo{author}{{Popovi{\'c}}, L.{\v C}.}, \bibinfo{year}{2012}.
\newblock \bibinfo{title}{{3D} radiative transfer modelling of the dusty tori
  around active galactic nuclei as a clumpy two-phase medium}.
\newblock \bibinfo{journal}{\mnras} \bibinfo{volume}{420},
  \bibinfo{pages}{2756--2772}.
\newblock \URLprefix
  \url{http://mnras.oxfordjournals.org/cgi/doi/10.1111/j.1365-2966.2011.19775.x},
  \DOIprefix\doi{10.1111/j.1365-2966.2011.19775.x},
  \href{http://arxiv.org/abs/1109.1286}{\tt arXiv:1109.1286}.
\bibitem[{Stenholm(1994)}]{Stenholm_94_Silicate}
\bibinfo{author}{Stenholm, L.}, \bibinfo{year}{1994}.
\newblock \bibinfo{title}{Silicate dust discs as sources of the {AGN
  IR}-emission.}
\newblock \bibinfo{journal}{\aap} \bibinfo{volume}{290},
  \bibinfo{pages}{393--398}.
\newblock \URLprefix \url{http://adsabs.harvard.edu/abs/1994A%26A...290..393S}.
\bibitem[{{Sturm} et~al.(2005){Sturm}, {Schweitzer}, {Lutz}, {Contursi},
  {Genzel}, {Lehnert}, {Tacconi}, {Veilleux}, {Rupke}, {Kim}, {Sternberg},
  {Maoz}, {Lord}, {Mazzarella} and {Sanders}}]{Sturmetal2005}
\bibinfo{author}{{Sturm}, E.}, \bibinfo{author}{{Schweitzer}, M.},
  \bibinfo{author}{{Lutz}, D.}, \bibinfo{author}{{Contursi}, A.},
  \bibinfo{author}{{Genzel}, R.}, \bibinfo{author}{{Lehnert}, M.D.},
  \bibinfo{author}{{Tacconi}, L.J.}, \bibinfo{author}{{Veilleux}, S.},
  \bibinfo{author}{{Rupke}, D.S.}, \bibinfo{author}{{Kim}, D.C.},
  \bibinfo{author}{{Sternberg}, A.}, \bibinfo{author}{{Maoz}, D.},
  \bibinfo{author}{{Lord}, S.}, \bibinfo{author}{{Mazzarella}, J.},
  \bibinfo{author}{{Sanders}, D.B.}, \bibinfo{year}{2005}.
\newblock \bibinfo{title}{{Silicate Emissions in Active Galaxies: From LINERs
  to QSOs}}.
\newblock \bibinfo{journal}{\apjl} \bibinfo{volume}{629},
  \bibinfo{pages}{L21--L23}.
\newblock \DOIprefix\doi{10.1086/444359},
  \href{http://arxiv.org/abs/astro-ph/0506716}{\tt arXiv:astro-ph/0506716}.
\bibitem[{{Tielens}(1990)}]{Tielens1990}
\bibinfo{author}{{Tielens}, A.G.G.M.}, \bibinfo{year}{1990}.
\newblock \bibinfo{title}{{Towards a circumstellar silicate mineralogy}}, in:
  \bibinfo{editor}{{Mennessier}, M.O.}, \bibinfo{editor}{{Omont}, A.} (Eds.),
  \bibinfo{booktitle}{From Miras to Planetary Nebulae: Which Path for Stellar
  Evolution?}, pp. \bibinfo{pages}{186--200}.
\bibitem[{{van Bemmel} and {Dullemond}(2003)}]{vanBemmel_03_New}
\bibinfo{author}{{van Bemmel}, I.M.}, \bibinfo{author}{{Dullemond}, C.P.},
  \bibinfo{year}{2003}.
\newblock \bibinfo{title}{New radiative transfer models for obscuring tori in
  active galaxies}.
\newblock \bibinfo{journal}{\aap} \bibinfo{volume}{404},
  \bibinfo{pages}{1--19}.
\newblock \DOIprefix\doi{10.1051/0004-6361:20030427},
  \href{http://arxiv.org/abs/astro-ph/0303496}{\tt arXiv:astro-ph/0303496}.
\bibitem[{{Veilleux} et~al.(2009){Veilleux}, {Rupke}, {Kim}, {Genzel}, {Sturm},
  {Lutz}, {Contursi}, {Schweitzer}, {Tacconi}, {Netzer}, {Sternberg}, {Mihos},
  {Baker}, {Mazzarella}, {Lord}, {Sanders}, {Stockton}, {Joseph} and
  {Barnes}}]{Veilleuxetal2009}
\bibinfo{author}{{Veilleux}, S.}, \bibinfo{author}{{Rupke}, D.S.N.},
  \bibinfo{author}{{Kim}, D.C.}, \bibinfo{author}{{Genzel}, R.},
  \bibinfo{author}{{Sturm}, E.}, \bibinfo{author}{{Lutz}, D.},
  \bibinfo{author}{{Contursi}, A.}, \bibinfo{author}{{Schweitzer}, M.},
  \bibinfo{author}{{Tacconi}, L.J.}, \bibinfo{author}{{Netzer}, H.},
  \bibinfo{author}{{Sternberg}, A.}, \bibinfo{author}{{Mihos}, J.C.},
  \bibinfo{author}{{Baker}, A.J.}, \bibinfo{author}{{Mazzarella}, J.M.},
  \bibinfo{author}{{Lord}, S.}, \bibinfo{author}{{Sanders}, D.B.},
  \bibinfo{author}{{Stockton}, A.}, \bibinfo{author}{{Joseph}, R.D.},
  \bibinfo{author}{{Barnes}, J.E.}, \bibinfo{year}{2009}.
\newblock \bibinfo{title}{{Spitzer Quasar and Ulirg Evolution Study (QUEST).
  IV. Comparison of 1 Jy Ultraluminous Infrared Galaxies with Palomar-Green
  Quasars}}.
\newblock \bibinfo{journal}{\apjs} \bibinfo{volume}{182},
  \bibinfo{pages}{628--666}.
\newblock \DOIprefix\doi{10.1088/0067-0049/182/2/628},
  \href{http://arxiv.org/abs/0905.1577}{\tt arXiv:0905.1577}.
\bibitem[{{Vestergaard} and {Peterson}(2006)}]{VestergaardPeterson2006}
\bibinfo{author}{{Vestergaard}, M.}, \bibinfo{author}{{Peterson}, B.M.},
  \bibinfo{year}{2006}.
\newblock \bibinfo{title}{{Determining Central Black Hole Masses in Distant
  Active Galaxies and Quasars. II. Improved Optical and UV Scaling
  Relationships}}.
\newblock \bibinfo{journal}{\apj} \bibinfo{volume}{641},
  \bibinfo{pages}{689--709}.
\newblock \DOIprefix\doi{10.1086/500572},
  \href{http://arxiv.org/abs/astro-ph/0601303}{\tt arXiv:astro-ph/0601303}.
\bibitem[{{Xie} et~al.(2014){Xie}, {Hao} and {Li}}]{Xieetal2014}
\bibinfo{author}{{Xie}, Y.}, \bibinfo{author}{{Hao}, L.},
  \bibinfo{author}{{Li}, A.}, \bibinfo{year}{2014}.
\newblock \bibinfo{title}{{A Tale of Three Galaxies: Anomalous Dust Properties
  in IRAS F10398+1455, IRAS F21013-0739, and SDSS J0808+3948}}.
\newblock \bibinfo{journal}{\apjl} \bibinfo{volume}{794}, \bibinfo{pages}{L19}.
\newblock \DOIprefix\doi{10.1088/2041-8205/794/2/L19},
  \href{http://arxiv.org/abs/1407.0914}{\tt arXiv:1407.0914}.
\bibitem[{{Xie} et~al.(2015){Xie}, {Li}, {Hao} and {Nikutta}}]{Xie_15_Tale}
\bibinfo{author}{{Xie}, Y.}, \bibinfo{author}{{Li}, A.},
  \bibinfo{author}{{Hao}, L.}, \bibinfo{author}{{Nikutta}, R.},
  \bibinfo{year}{2015}.
\newblock \bibinfo{title}{A tale of three galaxies: Deciphering the infrared
  emission of the spectroscopically anomalous galaxies {IRAS} f10398+1455,
  {IRAS} f21013-0739, and {SDSS} j0808+3948}.
\newblock \bibinfo{journal}{\apj} \bibinfo{volume}{808}, \bibinfo{pages}{145}.
\newblock \DOIprefix\doi{10.1088/0004-637X/808/2/145},
  \href{http://arxiv.org/abs/1507.03280}{\tt arXiv:1507.03280}.

\end{thebibliography}
\end{document}